\documentclass[traditabstract]{aa}  %{emulateapj}  [,referee]
\usepackage{boldline} %
\usepackage{graphicx}
\usepackage[colorlinks=true,
            linkcolor=red,
            urlcolor=green,
            citecolor=blue]{hyperref}

\usepackage{txfonts}

\usepackage{makecell}
\usepackage{float}

\newcommand{\myemail}{antonello.calabro@inaf.it}

\usepackage{multirow}
\usepackage{booktabs,tabularx}

\begin{document}

%\setlength{\parskip}{0.1pt}


\title{The environmental dependence of the stellar and gas-phase mass-metallicity relation at $2<z<4$}

\author{A. Calabr{\`o}\inst{1} %ok
\and L. Guaita\inst{2} %ok
\and L. Pentericci\inst{1} %ok
\and F. Fontanot\inst{3,4} %ok
\and M. Castellano\inst{1} %ok
\and G. De Lucia\inst{3} %ok
\and T. Garofalo\inst{1} %ok
\and P. Santini\inst{1} %ok
\and F. Cullen\inst{5} %ok
\and A. Carnall\inst{5} %ok
\and B. Garilli\inst{6} %ok
\and M. Talia\inst{7} %ok
\and G. Cresci\inst{8} %ok
\and M. Franco\inst{9} %ok
\and J.P.U. Fynbo\inst{11} %ok
\and N. P. Hathi\inst{12} %ok
\and M. Hirschmann\inst{3,10} %ok
\and A. Koekemoer\inst{12} %ok
\and M. Llerena\inst{13} %ok
\and L. Xie\inst{14} %ok
}

% A. Calabro, L. Guaita, L. Pentericci, F. Fontanot, M. Castellano, G. De Lucia, T. Garofalo, P. Santini, F. Cullen, A. Carnall, B. Garilli, M. Talia, G. Cresci, M. Franco, J.P.U. Fynbo, N. P. Hathi, M. Hirschmann, A. Koekemoer, M. Llerena, and L. Xie

% A. Calabrò, M. Castellano, L. Pentericci, F.Fontanot, N.Menci, F.Cullen, R.McLure, M.Bolzonella, A.Cimatti, F.Marchi, M.Talia, R.Amorin, G.Cresci, G.De Lucia, J.Fynbo, A.Fontana, M.Franco, N.P.Hathi, P.Hibon, M.Hirschmann, F.Mannucci, P.Santini, A.Saxena, D.Schaerer, L.Xie, G.Zamorani

\institute{INAF - Osservatorio Astronomico di Roma, via di Frascati 33, 00078, Monte Porzio Catone, Italy (\myemail)
\and Departamento de Ciencias Fisicas, Universidad Andres Bello, Fernandez Concha 700, Las Condes, Santiago, Chile
\and INAF - Astronomical Observatory of Trieste, via G.B. Tiepolo 11, I-34143 Trieste, Italy
\and IFPU - Institute for Fundamental Physics of the Universe, via Beirut 2, 34151, Trieste, Italy
\and SUPA, Institute for Astronomy, University of Edinburgh, Royal Observatory, Edinburgh EH9 3HJ % \thanks{Scottish Universities Physics Alliance}
\and INAF - IASF Milano, Via A. Corti 12, I-20133, Milano, Italy
\and INAF - Osservatorio Astronomico di Bologna, via P. Gobetti 93/3, I-40129, Bologna, Italy
\and INAF - Osservatorio Astrofisico di Arcetri, Largo E. Fermi 5, I-50125, Firenze, Italy
%\and University of Bologna, Department of Physics and Astronomy (DIFA) Via Gobetti 93/2- 40129, Bologna, Italy
%\and Instituto de Investigaci\'on Multidisciplinar en Ciencia y Tecnolog\'ia, Universidad de La Serena, Ra\'ul Bitr\'an 1305, La Serena, Chile
\and Centre for Astrophysics Research, University of Hertfordshire, Hatfield, AL10 9AB, UK
\and DARK, Niels Bohr Institute, University of Copenhagen, Jagtvej 128, 2100 Copenhagen, Denmark % MH
\and Cosmic Dawn Center, Niels Bohr Institute, Jagtvej 128, 2100, Copenhagen \O, Denmark
\and Space Telescope Science Institute, 3700 San Martin Drive, Baltimore, MD 21218, USA
%\and European Southern Observatory (ESO) Chile
\and Departamento de F\'isica y Astronom\'ia, Universidad de La Serena, Av. Juan Cisternas 1200 Norte, La Serena, Chile
%\and Department of Physics and Astronomy, University College London, Gower Street, London WC1E 6BT, UK
%\and Observatoire de Gen\`eve, Universit\'e de Gen\`eve, 51 Ch. des Maillettes, 1290 Versoix, Switzerland
%\and CNRS, IRAP, 14 Avenue E. Belin, 31400 Toulouse, France
\and Tianjin Astrophysics Center, Tianjin Normal University, Binshuixidao 393, 300384, Tianjin, China
}
\date{Received XXX}
%\and 18, INAF-Osservatorio Astronomico di Roma, via di Frascati 33, 00078, Monte Porzio Catone, Italy

\abstract 
{
In the local universe, galaxies in clusters typically show different physical and chemical properties compared to more isolated systems. Understanding how this difference originates and whether it is already in place at high redshift is still a matter of debate. Thanks to uniquely deep optical spectra available from the VANDELS survey, we investigate environmental effects on the stellar mass-metallicity relation (MZR) for a sample of nearly $1000$ star-forming galaxies in the redshift range $2<z<4$. We complement our dataset with MOSFIRE follow-up of $21$ galaxies to study the environmental dependence of the gas-phase MZR. Robust stellar and gas-phase metallicities are derived, respectively, from well-calibrated photospheric absorptions features at $1501$ and $1719$ \AA\ in the stacked spectra, and from optical emission lines ([OII]$\lambda \lambda 3726$-$3729$, [OIII]$\lambda5007$, and H$\beta$) in individual systems. We characterize the environment through multiple criteria by using the local galaxy density maps derived in the VANDELS fields to identify overdense structures and protoclusters with varying sizes.
We find that environmental effects are weak at redshifts $2<z<4$, and they are more important around the densest overdensity structures and protoclusters, where galaxies have a lower stellar metallicity (by $\sim 0.2$ dex) and a lower gas-phase metallicity (by $0.1$ dex) compared to the field, with a significance of $1\sigma$ and $2\sigma$, respectively. Crucially, this downward offset cannot be explained by a selection effect due to a higher SFR, a fainter UV continuum, or different dust attenuations and stellar ages for galaxies in overdense enviroments with respect to the field. 
In spite of the still low S/N of our results, we speculate about possible explanations of this environmental dependence. We propose a combination of increased mergers and high-speed encounters, more efficient AGN feedback in dense cores, and cold gas inflows from the cosmic web as viable physical mechanisms diluting the metal content of the cold gas reservoirs of overdense galaxies or expelling their metals to the intergalactic medium, even though additional studies are needed to conclude on the most influencing scenario. 
Finally, some tensions remain between observations and both semi-analytic models and hydrodynamical simulations, which predict no significant metallicity offset as a function of host halo mass, suggesting that an explicit implementation of environmental processes in dense protocluster cores is needed.
}

% Accepted for publication in A&A ; 22 pages, 15 figures, 2 tables, and 1 Appendix


\titlerunning{\footnotesize environmental dependence of the MZR}
\authorrunning{A.Calabr\`o et al.}
 \maketitle
\section{Introduction}\label{introduction}

Galaxies are not isolated systems, but during their lives they exchange baryonic matter with their gaseous haloes, known to as their circumgalactic medium (CGM) and, to a larger extent, with the intergalactic medium (IGM) pervading the space between them. These gas flows occurring at kiloparsec (kpc) scales or less are primarily responsible for their star-formation activity, mass growth, and metal enrichment across cosmic time \citep{peeples19}.

In addition to this scenario, also the large scale distribution of galaxies on megaparsec (Mpc) scales plays a major influence on their evolution. The extent to which this is important, and the impact on galaxy properties that reside in different environments has been object of intense debate in the last decades \citep{delucia04a,gao05,maulbetsch07,shankar13}. We know that denser environments at low redshift (e.g., mature galaxy clusters) host larger fractions of red early-type galaxies %, pag.66 MBW % Notably, galaxy evolution progresses faster in regions with a higher mass concentration
compared to the field, which translates into the so-called morphology - density relation \citep{dressler80}. On average, galaxies in overdense regions are also more massive, redder, less gas-rich, and less star-forming compared to their more isolated counterparts \citep{kauffmann04,baldry06,weinmann06}. 
This is due to a combination of mechanisms that come at play when galaxies approach a higher density region, and that are able to reduce or even completely quench their star-formation activity. These processes include harassment, dynamical friction and merging, ram-pressure stripping and starvation. Harassment is due to the interaction of galaxies with close companions and with the gravitational potential of the cluster, which can destroy galactic disks and produce more spheroidal shapes \citep{farouki81,moore96}. Dynamical friction and merging are responsible for the gradual loss of kinetic energy by the galaxies, which may approach the bottom of the gravitational potential well and merge together, forming the so-called Brighest Cluster Galaxy (BCG), the most luminous and massive system in an overdense structure \citep{ostriker75,delucia07}. Finally, ram pressure stripping refers to the removal of cold gas in the galaxies due to their interaction with the hot, X-ray bright, intracluster medium \citep[ICM,][]{gunn72}. If only the outer parts of the galactic disk are stripped, the star-formation is not quenched abruptly but continues over a longer period until the complete exhaustion of the available gas. In this case, the physical phenomenon is dubbed strangulation \citep{larson80,balogh00}. The consequences of ram pressure gas stripping are frequently seen in the local Universe in the so-called Jellyfish galaxies \citep{poggianti17}, and they can affect both the hot ionized gas \citep{jachym14} and the cold molecular gas \citep{poggianti19}, depriving galaxies from their fuel. 
% G.D.L. There's some confusion here as for which gas is removed: when people talk of ram-pressure stripping, they usually refer to the cold gas in the disk. Albeit the process is the same, starvation is used to refer to the removal of the hot gas reservoir associated with galaxies. At least I think this is the way it was intended in the original work by Larson. 
% G.D.L. In general, is the situation so clearcut in the local Universe? I have not been following this literature in detail but my impression was that, when controlling for stellar mass, environmental dependencies of metal content are generally very small.

Apart from well-studied individual systems in the local Universe, it is interesting to check how these processes impact the galaxy population as a whole. For this statistical approach, we need to observe the largest possible number of galaxies probing different regimes of environmental conditions, stellar masses, star-formation activity, and redshift. The environment is typically defined by deriving a three-dimensional local galaxy density map in the entire survey volume, from which connected overdense regions and clusters can be identified. 
For galaxies inside clusters, a useful distinction is made between centrals and satellites, where the former are sitting at the bottom of the dark-matter halo gravitational potential well and usually correspond to the BCG, while the remaining galaxies are called satellites. These two types of systems are thought to have evolved in a different way across cosmic time.
%In order  Then, galaxies in clusters can be classified according to their distance from the cluster centers or to their stellar mass.  
%It is relatively easier to perform this analysis in the local Universe, where the SDSS survey has mapped millions of galaxies ADD DETAILS (check websites ...) in different environments.
%Observationally, galaxies have been classified in different ways according to their environment.  
%Alternatively, we can also calculate a 3-dimensional galaxy density map in the whole field and associate to each object the density field corresponding to its spatial potition. This density increases naturally in cluster environments, while it decreases in the voids. Finally, once we define a cluster definition criterion, we can flag all the objects belonging to the same structure.
%Once the center of an overdensity is identified, it is also possible to measure and classify its members according to their radial distance from this center. 
%This paragraph is confusing. It is not particularly clear how to distinguish between central and satellites in observational samples. I guess here you are referring explicitly to clusters (where the BCG can be considered as the bonafide central galaxy), but you define this enviroment only later in the paragraph. Here you are still referring to general environments/

Observational studies based on the SDSS, the widest survey in the local Universe mapping one-quarter of the entire sky, have found that satellite galaxies with stellar masses M$_\odot < 10^{10}$ M$_\odot$ have enhanced gas-phase and stellar metallicity compared to equally massive centrals \citep{pasquali10,shimakawa15,bahe16}. 
In addition, star-forming satellites at all stellar masses show a clear correlation between their local density and their gas-phase metallicity, with the metal content increasing when we move from low to high density regions, but this behaviour is not observed for star-forming centrals \citep{peng14,maier19}.
Regarding the stellar phase mass-metallicity relation (MZR), no environmental effects are found for star-forming galaxies as a function of local density and distance from the cluster centers, while a small systematic offset of $<0.05$ dex is only observed for passive and green valley systems \citep{trussler21}.

Broadening the knowledge to how the large scale structures affect galaxy properties at earlier cosmic epochs is essential. Galaxy clusters are thought to be the earliest fingerprint of galaxy formation, and they are crucial to understand the processes responsible for their evolution, in particular for the triggering and suppression of star-formation and black-hole activity \citep{kravtsov12}.
Despite their importance, environmental effects are still poorly understood at high redshifts. The cosmic epoch between $z=1$ and $z=2$ seems to be crucial for the modification of fundamental properties of galaxy clusters. For example, \citet{elbaz07} claim that the star-formation - density relation reverses at $z\sim1$, with individual SFRs of galaxies at $z>1$ increasing with local galaxy density. Most galaxy clusters at that early epochs were not completely virialized and were still in their process of formation, with larger sizes and more loosely distributed members. For this reason, they are often referred to as protoclusters, lacking both of a well-defined red sequence galaxy population \citep{salimbeni09} and of a hot intracluster medium (ICM) that we commonly observe in X-rays in mature clusters at $z\sim0$ \citep{overzier16}. 

At high redshift, the investigation of the environmental influence on the metal content of galaxies is usually limited to their gas-phase metallicity, which is easier to measure from bright optical and UV emission lines. Despite this, several studies have found completely different results so far. First, \citet{tran15}, \citet{kacprzak15}, and \citet{namiki19} claim that there is very small and no significant impact of the environment on the MZR at redshifts $1.5 < z < 2$. \citet{kulas13} and \citet{shimakawa15} found instead a significantly higher gas-phase metallicity in cluster galaxies, which they interpreted as due to metal enriched outflowing gas falling back onto the galactic disk on short timescales because of the higher pressure of intergalactic medium (IGM) in dense regions. Finally, \citet{valentino15} showed that cluster members at $z\sim 2$ could be more metal poor than equally massive field galaxies, because of metal dilution driven by pristine gas inflows from the cosmic web, which are more efficient at earlier cosmic epochs. 
%Ad alto redshift c'e' la reversal della SFR-density e morpho-density anche ?) relation (vedi Elbaz et al. 2007), inoltre le condizioni dell'Universo sono differenti. Infatti molti cluster come li vediamo oggi non si sono ancora virializzati, per cui spesso si parla di protoclusters, ovvero strutture in formazione che seguono l'assemblaggio e il collasso degli aloni di materia oscura. In queste strutture, le galassie all'interno sono ancora molto sparse. 
% however, such studies at high-redshift are more difficult to perform and are usually limited to the gas-phase metallicity, hence the picture is still far from being clear. Besides this, it is debated which are the environmental effects on large scales playing a major role on galaxy evolution. Different works have indeed found different results : Vedi Namiki che fa un bel sunto ... 1,2, 3.
% DICI anche perche' e' importane studiare environmental effects e clusters ad alto redshift. 
Interestigly, a recent study by \citet{chartab21} seems to support this latter scenario, claiming that the environmental density dependence of the gas-phase MZR reverses at around redshift $2$. According to them, HII regions in overdense regions at $z>2$ are more metal poor than those in the field at similar stellar masses, as due to more efficient gas cooling during the early phases of cluster formation, hence to a higher fraction of metal-poor gas in these systems with respect to $z\sim 0$. 
Overall, these studies suggest that different mechanisms might be at play at the same time, and which can affect the metal content and other physical properties of galaxies in overdense regions. While a larger statistics is needed to clarify the dominant scenario, we still do not have any knowledge about the effects of the environment on the stellar metallicity at these same redshifts. 

\begin{figure*}[h!]
    \centering
    \includegraphics[angle=0,width=\linewidth,trim={1cm 0.cm 2cm 1.5cm},clip]{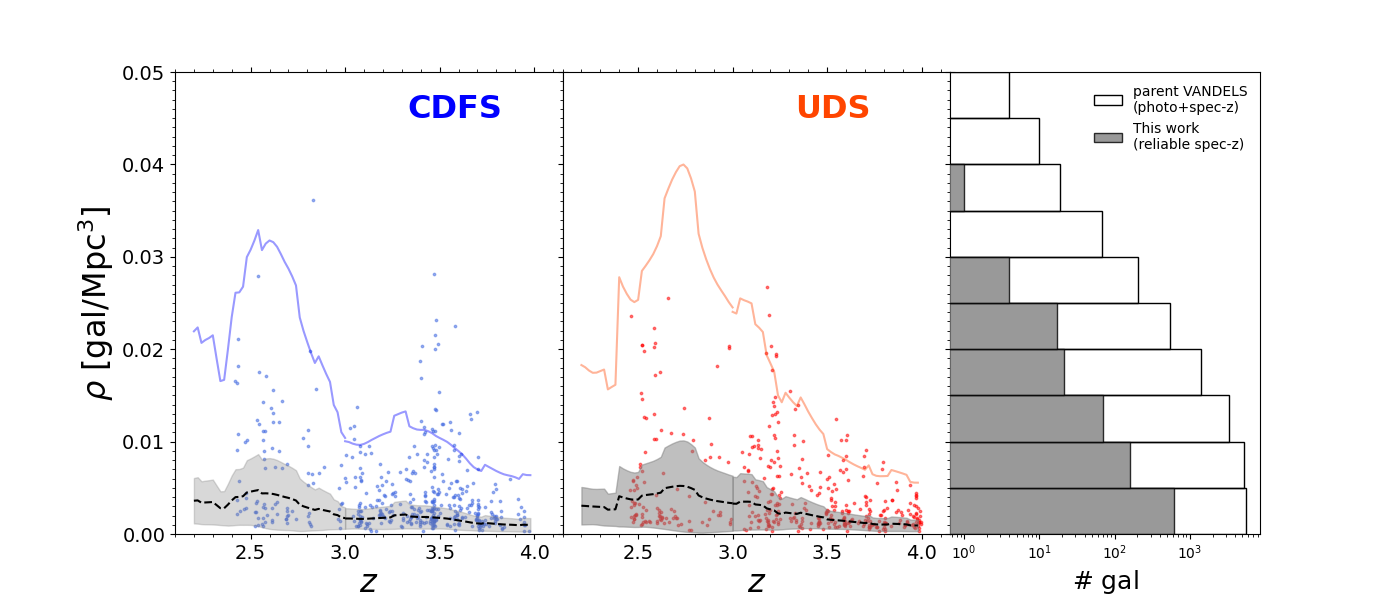}
    \vspace{-5mm} 
    \caption{\small \textit{Left and middle:} Redshift ($z$) vs local galaxy density ($\rho$) for the galaxies with reliable spectroscopic redshifts ($z$-flag $\geq3$) that we use to test the environmental effects on the MZR, in the CDFS and UDS fields, respectively with blue and orange dots. We remark that these are only a subset of the parent VANDELS sample used to calculate the galaxy density maps. The black dashed line and the gray shaded region represent for each $z$, the median local density $\rho_{med} (z)$ and the corresponding standard deviation, respectively. The colored continuous lines highlight instead the criterion adopted to consider galaxies as members of an overdensity structure, that is, $>6 \sigma$ above the median density. \textit{Right:} Histogram distribution of the local galaxy density for the VANDELS sample used in this work (both fields). The white empty bars show the local density distribution of the parent VANDELS sample (having both photometric and spectroscopic $z$) used for the calculation of the density and for the identification of overdense structures. %In all panels, the horizontal green dotted line represents the median local density of the reliable spectroscopic sample, which is adopted to separate the sample between lower and higher density regions in Section \ref{results}.
    }\label{density_histogram}
\end{figure*}

In this paper, we investigate for the first time at redshift $>2$ the environmental dependence on the stellar mass - stellar metallicity relation, which, compared to the gaseous phase, provides an important and independent diagnostic of the metal content and evolution in galaxies from $10$ to $12$ billion years ago. This is possible thanks to the uniquely deep spectra provided by the VANDELS survey \citep{pentericci18,mclure18,garilli21} for thousand of galaxies, probing the bulk of the star-forming galaxy population at this early cosmic epoch. The high S/N reached for the stellar continua ($>7$ per resolution element for $80\%$ of the observed sample) allow indeed accurate estimates of the metal content in stars from spectral stacks, as showed by \citet{cullen19} and \citet[][hereafter named C21]{calabro21}. In addition, we also provide further constraints to environmental driven variations of the gas-phase metallicity, exploiting a recent near-infrared follow-up for a subset of VANDELS targets \citep{cullen21}. 
For the analysis of the environment, we build upon the results of \citet[][hereafter named G20]{guaita20}, who identified protocluster structures between $z=2$ and $z=4$ from the 3D galaxy density map in the cosmic volume targeted by VANDELS, reconstructed through the (2+1)D cluster finding algorithm of \citet{trevese07} and \citet{castellano07}. 

The paper is organized as follows. In Section~2, we describe the photometric parent sample from which we derive the galaxy density field in the VANDELS area and identify overdensity structures and protoclusters with different approaches. Then, we introduce the final spectroscopic sample and illustrate the derivation of the stellar and gas-phase metallicity, for stacked spectra and individual systems, respectively. 
We present the results of these measurements in Section~3, showing the environmental dependence of the mass-metallicity relation, both as a function of the galaxy density and protocluster or overdensity membership. 
Finally, in Section~4, we discuss the physical scenarios that are able to explain our observed trends, and compare our results to the predictions of different semi-analytic models of galaxy evolution and hydrodynamical simulations. A summary with the main conclusions is featured in Section~5. 
In our analysis, we adopt a \citet{chabrier03} initial mass function (IMF), a solar metallicity Z$_\odot$ $=0.0142$ \citep{asplund09} and, unless otherwise stated, we assume a cosmology with $H_{0}=70$ $\rm km\ s^{-1}Mpc^{-1}$, $\Omega_{\rm m} = 0.3$, $\Omega_\Lambda = 0.7$. %We also assume, by convention, a positive equivalent width (EW) for absorption lines and a negative EW for lines in emission.

\section{Methodology}\label{methodology}

In this section, we briefly present our parent catalog and the subset targeted by the VANDELS survey. Regarding the former, we show the derivation of the photometric redshifts and other physical properties (e.g., stellar masses, SFR) obtained from their optical to infrared SEDs. We also describe the method adopted for the estimation of the local galaxy density in the whole VANDELS fields, and for the identification of overdense structures at redshifts $2 < z < 4$. 
Then, we present the selection of the final sample for our analysis and show how the stellar metallicity is derived from spectral stacks of galaxies in bins of different stellar masses. Finally, we introduce a smaller sample of VANDELS galaxies that were followed up in the near-infrared, and for which individual estimates of the gas-phase metallicity are derived.

%\subsection{VANDELS observations (NEW)}\label{observations}

\subsection{The photometric parent sample and SED fitting with Beagle}\label{parent}

\begin{figure*}[ht!]
    \centering
    \includegraphics[angle=0,width=0.9\linewidth,trim={3cm 0.cm 3cm 1.cm},clip]{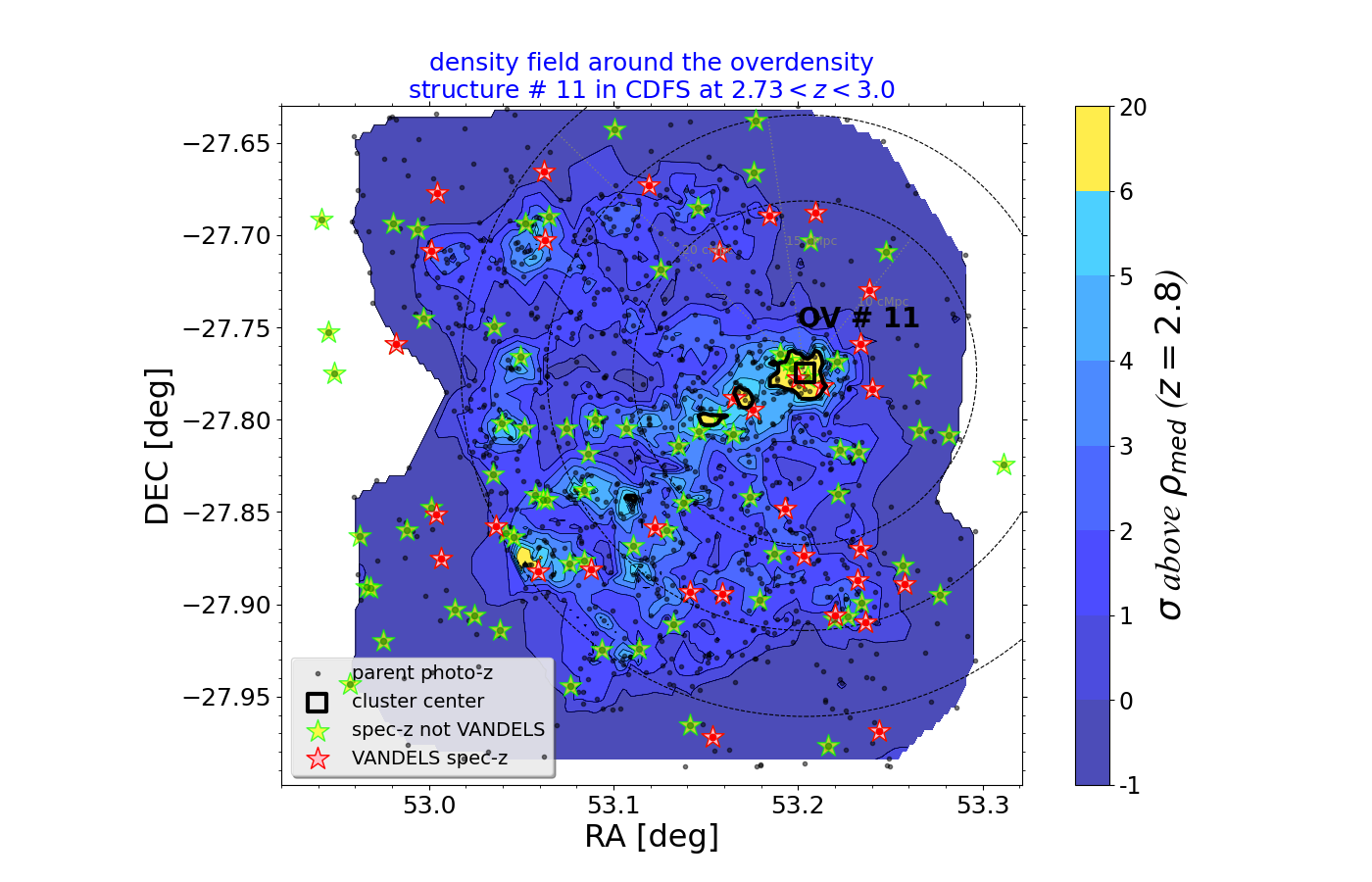}
    \vspace{-5mm} 
    \caption{\small Example of an overdensity structure identified at $z \simeq 2.8$ in the CDFS field. We show overplotted the contours of the local galaxy density (gal/Mpc$^3$) estimated with the method of \citet{guaita20}, with different levels corresponding to the enhancement in units of $1\sigma$ compared to the average galaxy density and standard deviation in the given redshift range. The symbols represent the galaxy sample used for the determination of the density map in Section \ref{densityfield}. In particular, the black points, the red stars and the green stars identify, respectively, galaxies with photometric redshifts, the VANDELS sample with spectroscopic redshift flag $\geq2$, and galaxies with spectroscopic redshift coming from other surveys than VANDELS. The big empty black square and the continuous black line represent, respectively, the center of the overdensity structure and its boundaries in projected space. The black dashed circles with increasing radii of $10$, $15$, and $20$ cMpc from the overdensity structure centers are used for an alternative definition of protoclusters, assuming a $\Delta z=0.04$ along the redshift direction. We remind that only VANDELS galaxies with $z$-flag $\geq3$ are used for the stacking and metallicity estimations in Section \ref{stellar-metallicity}.  % Poi mettere anche il campione spettroscopico globale NON VANDELS con delle stelle diverse. 
    }\label{map1}
\end{figure*}

The galaxies considered in this work have been originally selected for the VANDELS survey, a uniquely deep spectrocopic survey with the VIMOS multi-object spectrograph at the ESO-VLT \citep{pentericci18,mclure18,garilli21}. This survey mainly targeted star-forming galaxies at redshifts $2$-$6.5$ in the Ultra Deep Survey (UDS) and Chandra Deep Field South (CDFS) fields, which cover together an area of the sky of $\sim 0.2$ deg$^2$. Nearly half of this original sample falls within the CANDELS region \citep{grogin11,koekemoer11} %\citep{grogin11,koekemoer11}
, for which deep H-band-selected photometric catalogs are already available in both CDFS and UDS, reaching $\sim90 \%$ completeness at H$_{AB}$ $\sim25.5$ and $\sim26.7$, respectively \citep{guo13,galametz13}. The remaining galaxies are distributed instead over a wider region, and new photometric catalogs are generated for them within our collaboration by using the publicly available imaging in optical and near-infrared (NIR) bands, with a depth of $H_{AB}=$ $24.5$ ($25$) in the CDFS (UDS) fields. 

For the VANDELS parent sample, both with HST and ground-only observations, we carry out an SED fitting to determine the basic physical properties of the galaxies, such as stellar mass, SFR, and stellar age. We used the software Beagle, acronym for BayEsian Analysis of GaLaxy sEds \citep{chevallard16}, which is a very powerful and versatile tool that allows the user to simulate the full UV to IR theoretical spectral energy distribution of a galaxy. Beagle's SED fitting algorithm is based on Bayesian inference, which iteratively compares theoretical SEDs to the observations according to the Multinest algorithm \citep{feroz09} to find the best-fit parameters. 

In Beagle, both the stellar and nebular emission components of the galaxy are considered. Regarding the former, we adopt \citet{gutkin16} stellar population models with a Chabrier IMF. The star-formation history (SFH) is parametrized using an exponentially delayed analytical form as $\tau e^{- \tau}$, where $\tau$ is the typical star-formation time. The star-formation rate timescale, which is the final time bin where the SFR of the galaxy is considered constant and calculated, is set to $100$ Myr. 
In our case, we assume a uniform prior for $\log \tau$ in the range $7.0$-$10.5$ (in Myr), and for the specific SFR (SSFR) in the range $-9.5 < \log$(SSFR) $< -7.05$ (in yr$^{-1}$). We also set the maximum stellar age, defined as the age of the oldest stellar population in the galaxy, with a uniform prior in the logarithmic range $7.0$-$10.2$ (yr). 
Finally, the stellar mass M$_\star$ and the stellar metallicity Z$_\star$, both outputs of the program, are initialised, respectively, with a uniform prior in the range $5 < \log (M_\star/M_\odot) < 12$ and with a gaussian prior centered in $ \log (Z_\star/Z_\odot) = -0.85$ with $1 \sigma=0.2$, corresponding to the average stellar metallicity of VANDELS galaxies \citep{cullen19,calabro21}. The redshifts of the galaxies are fixed to the spectroscopic value when available (see the following section). 

The nebular component is modeled through the photoionization models of \citet{gutkin16}, which combine the above stellar population models with the photoionization code CLOUDY v.13.3 \citep{ferland13}. 
The HII regions and diffuse ionized gas are thus described through some effective parameters representative of the whole galaxy, including the ISM metallicity Z$_\text{ISM}$ and the ionization parameter log(U). While the first is dependent on other quantities, the latter is free to vary in the range $-4 < \log (U) < -1$ with a uniform prior.  
Regarding the dust, we adopt a \citet{charlot00} attenuation law and we assign a uniform prior to the V-band optical attenuation depth $\tau_V$ in the range 0-10 mag. Finally, the dust-to-metal ratio $ \xi_{d}$ and the ratio between interstellar medium depth and total optical depth $\mu = \tau^{ISM}/ \tau_V$, are both set to their default values of $0.3$, consistent with standard values reported in the literature \citep{camps16,battisti20}. With these parameters, the intensity of emission lines generated by simple stellar populations with age $<10^8$ yr are calculated and taken into account by Beagle in the SED fitting procedure.
The \citet{inoue14} model is also included to describe the inter-galactic medium (IGM) transmission.  
% la metallicità dell’ISM ZISM = 0.0001 - 0.04 (che corrisponde ad un ratio F e/H = −2.1 - 0.42) - QUINDI SICURO DI QUESTA COSA ? 
We remark that the stellar masses are very stable against the choice of the SFH and dust attenuation law, as shown by \citet{santini15}, and they have typical statistical uncertainties of $0.1$ dex. They also agree well with the values obtained through \textsc{Bagpipes} \citep{carnall18} and adopted in the final DR4 VANDELS release \citep{garilli21}. 

\subsection{Galaxy density map in VANDELS fields}\label{densityfield}

\begin{figure*}[h!]
    \centering
    \includegraphics[angle=0,width=0.9\linewidth,trim={1cm 0.cm 2cm 1.cm},clip]{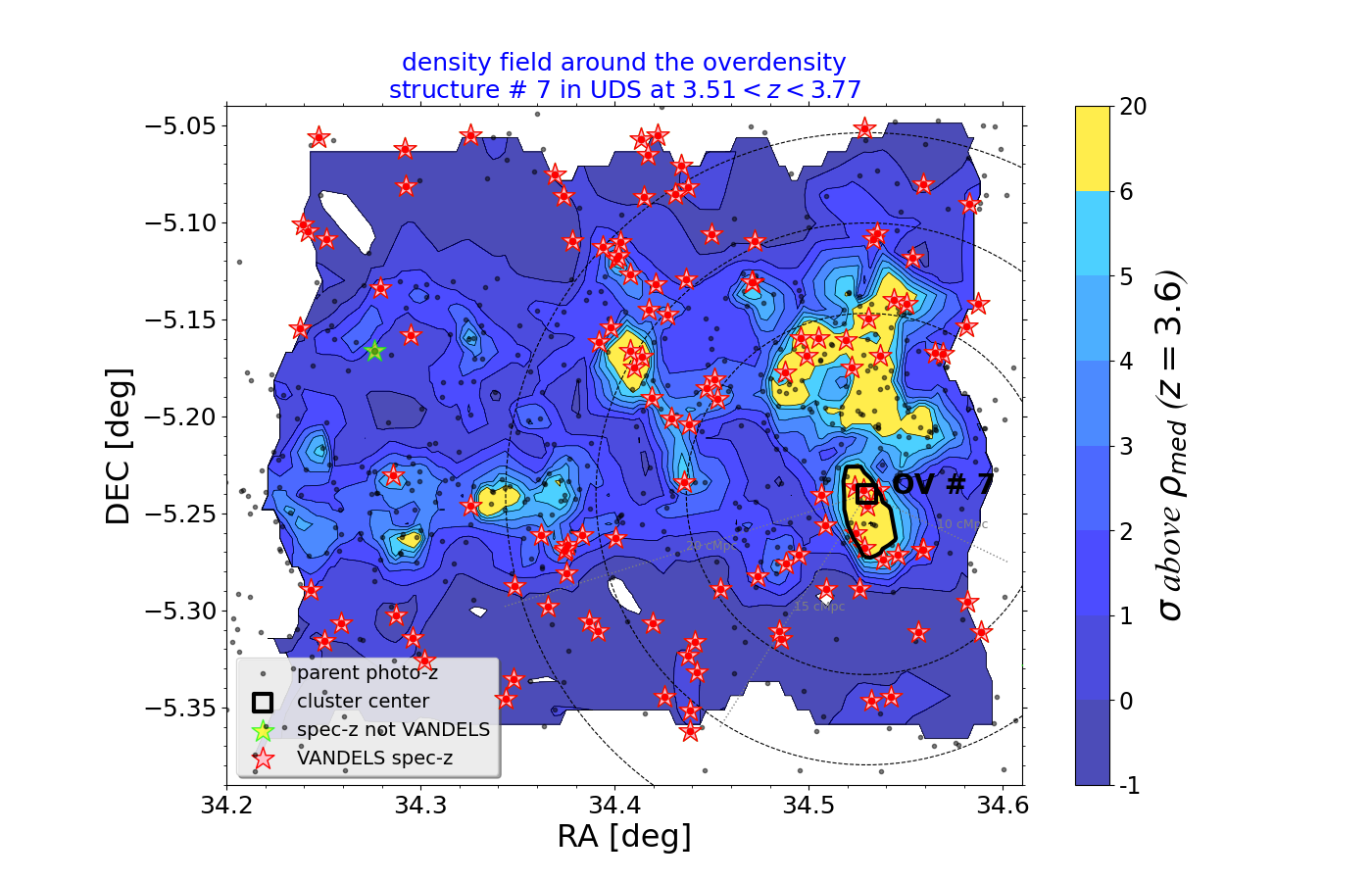}
    \vspace{-5mm} 
    \caption{\small Example of an overdensity structure identified at higher redshift ($z \simeq 3.6$) in the UDS field. Symbols and contours are the same as in Fig. \ref{map1}. For this particular overdensity structure at high-$z$, most of the spectroscopic redshifts come from the VANDELS survey.
    }\label{map2}
\end{figure*}

%The photometric parent sample of VANDELS will be used to calculate the galaxy density field in the whole VANDELS fields in Section \ref{densityfield}.
The environmental density $\rho_{gal}$, or otherwise said the local galaxy density field, is defined as the number of galaxies per Mpc$^3$. We calculate this quantity in the wide VANDELS area (CANDELS + GROUND regions) as described in G20. We summarize here the main features of this method, while referring to the previous paper for more details. 
The estimation of $\rho_{gal}$ is based on the (2+1)D algorithm developed by \citet{trevese07} and \citet{castellano07}, and used in several works, including \cite{salimbeni09}, \citet{castellano11}, \citet{pentericci13}, and \citet{galametz13b}. This algorithm considers as an input the 3D coordinates of all the objects in the catalog, namely the right ascension (RA), declination (DEC) in the sky, and the redshift. We use the spectroscopic redshift of the objects when available. 
For VANDELS galaxies, we consider those with spectroscopic redshift flag $3$, $4$, and $9$, which correspond to spectra with highly reliable redshift measurements, that is, $>95\%$ probability of being correct. We also include galaxies with spectroscopic flag $=2$ (i.e., $>75\%$ probability of having a correct redshift estimation) in order to improve the accuracy of the local density maps\footnote{We consider complementary catalogs to extend the number of spectroscopic redshifts available in the CDFS and UDS fields, allowing more accurate estimations of the galaxy density field. In CDFS, we adopt a supplementary spectroscopic redshift compilation made by Nimish Hathi (private communication), while in UDS we consider the results of mutiple surveys that will be collected in Maltby et al. (in prep). More details on the additional spectroscopic dataset are given in G20.}. For all the other systems, we adopt their photometric redshift estimate. This method is applied in the redshift range between $2$ and $4$, where $2$ is the lower $z$ limit of the survey, while the upper cut is set to mitigate the incompleteness of the photometric catalog at very high-$z$, and because the number of spectroscopic sources drops at $z>4$, mining the reliability of the results. These conditions yield 1311 (1019) $z_{spec}$ and 6235 (8531) $z_{phot}$ sources in the CDFS (UDS) fields, which are then used for this work. 
We note that we consider the complete VANDELS data release \citep{garilli21}, reaching an even larger fraction of spectroscopically confirmed redshifts (with z-flag $\geq 2$) compared to the previous calculations of G20. Even though this improves the local density reliability for the additional spectroscopic subset, it does not significantly affect the global overdensity structures found in G20.

The algorithm divides the entire volume survey in small cells with size $\delta$ RA $=3''$, $\delta$ DEC $=3''$, $\delta$ z $=0.02$, where the angular distance corresponds to $\sim75 $ comoving kilo-parsec (ckpc) at $z\sim2$, while the cell size in the redshift direction was set from simulations and was chosen as compromise between the spectroscopic and the photometric redshift accuracy \citep{castellano07}. Running the code on a mock galaxy catalog, G20 demonstrates that, with a $\delta z =0.02$, the algorithm is able to recover all the highest density peaks of simulated overdensities.
Then, for each object, the local galaxy density is defined as $\rho_N = N/V_N$ (hereafter simply $\rho$), where V$_N$ is the comoving volume which includes the N nearest neighbours, with N fixed to $10$. % Using 10 to 20 does not change the results.
In case ten counts are not reached, a maximum volume is considered for the calculation of $\rho_N$, which has an extension of $15''$ in the spatial direction and $0.04 \times (1+z)$ in the redshift direction. 
The distributions of the local galaxy densities $\rho$ for the parent photometric and the VANDELS spectroscopic subset in CDFS and UDS fields are presented in Fig. \ref{density_histogram}, also as a function of redshift within the range adopted here.

%\subsection{Identification of overdensities and protoclusters}\label{densityfield}

The algorithm is able to identify galaxy overdensities by comparing the local density value $\rho_{gal}$ in each volume cell to the mean density $\rho_{gal,med}(z)$ at a given redshift, and then extracting connected regions where $\rho_{gal}$ is at least $6 \sigma$ above the median value and contains a minimum number of ten cells and five galaxies. The RA-DEC-$z$ coordinates of the highest S/N density peak in the same structure is taken as the protocluster center. Among the identified overdensities, only those containing dense cores are kept. Dense cores at $z<3$ and $>3$ are characterized by more stringent conditions, that is, at least five galaxies located within a maximum projected radius of $7$ cMpc in the lower-$z$ bin and three galaxies within $8$ cMpc in the upper $z$-bin, where the radii are the typical protocluster sizes at this cosmic epoch \citep[e.g.,][]{springel05a,franck16}. The requirement of a dense core and the exact values of all the above parameters was widely tested with simulations in G20, and it was found to provide the most reliable structures. %Nevertheless, we remind that the final results would not be significantly affected by choosing slightly different settings for the identification of connected overdensities. 

This procedure yields $13$ structures in CDFS and $9$ structures in the UDS field in our redshift range (G20). In order to keep track of the membership, we also assign to each galaxy an `overdensity' flag (OV-flag, $=1$ or $0$), indicating, respectively, whether they belong or not to one of these overdensity structures. 
We remark that the majority of the identified overdensities are elongated along the redshift direction and composed by multiple density peaks at slightly different $z$. Usually, they are also formed by more than one peak in projected RA-DEC space, with a typical size of $1$-$4$ cMpc. As shown in G20, the overdensity members show rest-frame colors and specific SFRs typical of star-forming galaxies at $z<3$ and Lyman-break galaxies at $z>3$, suggesting that we are witnessing young protoclusters. Considering the mass density and the characteristic redshifts of the structures, they might evolve into a Fornax type virialized cluster by redshift $1$ to $0.2$.
In Fig. \ref{map1} and \ref{map2}, we show as an example two overdense structures, the first at $z<3$ in the CDFS field, while the second is at higher redshift and in UDS. The lower number of galaxies detected in the higher redshift bin follows the decrease in number density expected for our mass range between $z\sim2.5$ and $3.5$. 
We refer to G20 for a detailed description of the density distribution and galaxy map for all the remaining identified structures in VANDELS. 

In order to provide a more complete description of the galactic environment, we also define protocluster (PC) structures as cylindrical volumes around the overdensity centers identified above, with increasing radii of $10$, $15$, and $20$ cMpc, and redshift elongation within $\pm0.04$, corresponding to the average size in redshift space (i.e., $4000$ km/s, constant between $z=2$ and $4$) of all overdensities in VANDELS (G20). The lowest radius in projected space is the minimum which allows enough statistics in protocluster regions for our goals. The sky radiii that we inspect agree with the sizes of $10$-$20$ cMpc expected at $z>2$ for the progenitors of the present day clusters \citep{muldrew15,contini16}.
On the other hand, choosing radii larger than $20$ cMpc would produce an area that is comparable or larger than the whole CDFS or UDS field, and in most cases more than $50\%$ of its area would lie outside of the VANDELS region that we are analyzing. These protoclusters defined with different radii are drawn with black dashed circles in Fig. \ref{map1} and \ref{map2}. VANDELS galaxies (i.e., big green stars) belonging to them are assigned a protocluster flag (PC-flag$_\text{radius}$) of $1$, as opposed to external objects that have PC-flag$_\text{radius}=0$. 

We finally note that the different environmental definitions are not perfectly correlated, as, for example, galaxies outside of protocluster and overdensity structures do have a variety of density values, possibly including overdense systems that do not satisfy the conditions for a dense core, or are too far from them. On the other hand, our latter definition of protoclusters includes also systems that are locally more isolated, but that might be gravitationally influenced by the densest cores. 

\subsection{Spectroscopic sample and final selection}\label{spectroscopic}

In order to derive accurate estimates of the stellar metallicity to study the environmental dependence of the MZR, we need an extended and homogeneous spectroscopic dataset. From the parent catalog adopted for the local density field calculation in the previous section, we restrict now our analysis to galaxies, representing a fraction of $\sim10 \%$ of the initial sample, that have been selected in an unbiased way by VANDELS for spectroscopic follow-up. As discussed in previous works \citep{mclure18} and also shown later in this paper, this subset is fully representative of the Main Sequence of star-formation at redshift $\sim 3$ \citep{whitaker14,speagle14,schreiber15}, and has a specific SFR (SSFR) lower limit of $\sim 0.1$ Gyr$^{-1}$. 
For more details about the observations, full data reduction and redshift estimation, we refer to the two VANDELS introductory works and to the final release paper by \citet{garilli21}, which delivered to ESO $2087$ 1D calibrated spectra \footnote{All the catalogs and spectra are publicly available on the official VANDELS webpage \url{http://vandels.inaf.it}, and on the ESO Archive \url{https://www.eso.org/qi/}.}
We specify here that only galaxies with spectroscopic redshift flags $3$, $4$, and $9$ are considered in the stacking procedure and for the analysis of the stellar metallicities (hence excluding z-flag $2$ that have been used for the density map). 
These criteria yield in total $926$ galaxies, of which $476$ are in the CDFS field and $450$ lie in the UDS field. 

\subsection{Stellar metallicity estimates from stacked spectra}\label{stellar-metallicity}

We estimate the metallicity from the two stellar photospheric absorption features located around $1501$ and $1719$ \AA, following the methodology presented in C21. The first metallicity indicator, introduced by \citep{sommariva12}, is a single absorption line due to the SV ion, and arising in the photosphere of young, hot stars. The latter is instead a blend of multiple absorption lines, including NIV (1718.6 \AA), SiIV (1722.5, 1727.4 \AA), and multiple transitions of AlII and FeIV ranging from 1705 to 1729 \AA. Our metallicities are thus representative of the metal content of young stellar populations (O-B stars), which dominate the continuum emission in the FUV regime.  

The metallicity measurements are based on the equivalent width (EW) of the two absorption features, defined as $\int_{\lambda_1}^{\lambda_2} \left( f(\lambda)-f_{\rm cont}(\lambda) \right) /f_{\rm cont}(\lambda) d\lambda$, where $f(\lambda)$ is the spectrum (in erg/s/cm$^2$/\AA), $f_{\rm cont}(\lambda)$ is the pseudo-continuum flux, estimated through a linear interpolation of the average flux density between the closest blueward and redward pseudo-continuum windows defined by \citet{rix04}. The wavelengths $\lambda_1$ and $\lambda_2$ are the limits of the two features, that is, $1496$-$1506$ \AA\ for the former index, and $1705$-$1729$ \AA\ for the latter.

Finally, the stellar metallicities $\log_{10}$(Z$_\star$/Z$_\odot$) are derived from the EW of the individual indices applying the following third-order polynomial fits reported in C21:
\begin{equation}\label{calib1719}
\begin{aligned}
\mathrm{log}(Z/Z_\odot)|_{1501}=1.24 EW_{1501}^3-2.97 EW_{1501}^2+3.48 EW_{1501}-1.98 \\
\mathrm{log}(Z/Z_\odot)|_{1719}=0.06 EW_{1719}^3-0.44 EW_{1719}^2+1.51 EW_{1719}-2.12
\end{aligned}
\end{equation}
and then taking the mean of the two metallicity estimates. We remind, as shown in Fig.~3 of C21, that the two calibrations produce metallicities from the 1501 and 1719 indices that are consistent to each other within the errors. Therefore, the mean estimation has only the effect of decreasing the final associated uncertainty.

The calibrations are based, for consistency with the other VANDELS works, on Starburst99 stellar population models \citep{leitherer99}, which have an original resolution of $0.4$ \AA. These were resampled to match the wavelength grid used in our stacked spectra, and smoothed to the same average VANDELS resolution. In addition, the two equations \ref{calib1719} are derived assuming a Kroupa IMF (with upper mass cutoff of $100$ M$_\odot$) and a constant SFH of 100 Myr. Even though the individual galaxies can be younger than our adopted star-formation timescale, in spectral stacks we are averaging systems with different stellar ages in general, and $100$ Myr represents an average value that is representative of the stellar populations that mostly contribute to the FUV wavelength regime.
We also remark that, as shown in C21, the two indices considered here are the only good and reliable metallicity tracers in the far-UV range and at the resolution of VANDELS (R$=600$), as they are unaffected by ISM absorption \citet{vidalgarcia17} and largely independent on IMF, age, and dust attenuation. 
In particular, referring to C21, variations of the dust attenuation or a higher IMF cutoff mass of $300$ M$_\odot$ do not significantly affect the calibrations reported in equations \ref{calib1719}. On the other hand, a different IMF shape (e.g., Salpeter) and varying stellar ages in the range $50$ Myr - $2$ Gyr, would produce a variation in the derived metallicity of $\sim0.025$ dex for the $1501$ index and $<0.01$ dex for the $1719$ index (i.e., an average variation of $\lesssim0.015$ dex) in the metallicity regime probed by VANDELS, hence they would have a negligible impact on the results of this paper. As also shown in \citet{cullen19}, using instead a different set of models like BPASS (version 2.1) would yield lower metallicity estimates with a constant offset of $0.09$ dex, hence would not affect differences in the metal content between two galaxy populations, provided that the same type of models are adopted throughout the paper.
Finally, the statistical uncertainties on the EWs are instead determined from Monte Carlo simulations, perturbing the flux density at each spectral wavelength according to its $1\sigma$ error, and then propagating the results to $\log_{10}$ (Z$_\star$/Z$_\odot$).

Given the faintness (i.e., low absorption EW) of stellar photospheric features, the above method cannot be applied to single spectra, but stacking is required to enhance the S/N level up to values $\gtrsim 20$. This allows to properly constrain the final stellar metallicity within $0.1$-$0.2$ dex, as also found in C21 with the support of simulations.
In practice, rest-frame individual spectra, normalized to the wavelength range $1570$-$1601$ \AA\ (which is free of strong emission or absorption features and lying between our two metallicity indicators), are resampled to a common wavelength grid of $1$ \AA\ per pixel, and then stacked together taking the simple median. The noise composite spectrum is derived from bootstrap resampling. We note that simply applying the error propagation formula as in \citet{guaita15} we obtain a similar average noise level, hence it does not alter the uncertainties on the metallicity results. 
Overall, we refer to C21 for a more complete description of the metallicity derivation, for the analysis of other UV absorption line features as a function of IMF, age, and ISM contamination, and for the analytic relation between the SNR of the stellar continuum and the metallicity uncertainty achievable.
% We remind that this methodology provides results that are quantitatively in agreement (within $1 \sigma$) with those estimated through a dynamic nested sampling bayesian approach, based on a global fitting of the whole FUV spectrum, as performed by \citet{cullen19}. 

\subsection{Gas-phase metallicity from complementary near-infrared spectra}\label{gas-metallicity}

A small subset of VANDELS galaxies included in this work was followed up in the optical rest-frame with MOSFIRE in H and K band, and the results of this spectroscopic survey (dubbed NIRVANDELS) are presented in \citet{cullen21}.
This survey mainly observed star-forming galaxies in the CDFS and UDS fields, with H$_\text{AB} \leq 25.5$, and in the redshift range between $2.95$ and $3.8$, so that the brightest optical emission lines, ranging from [OII]$\lambda \lambda$ $3726$-$3729$ \AA\ to [OIII]$\lambda \lambda$ $4959$-$5007$ \AA, were all falling in nearly transparent atmospheric windows, either in band H or K. 

We use the metallicity measurements in the publicly available catalog of \citet{cullen21} to test the environmental dependence of the gas-phase metallicity at our redshifts. In this survey, gas-phase metallicity values are present for $21$ sources of the original VANDELS dataset. Thanks to the broad spectral coverage of MOSFIRE, it was possible to study the chemical abundance by combining the following emission-line ratios: [OIII]$5007$/[OII]$\lambda \lambda$ $3726$-$3729$ (O$_{32}$), [OIII]$5007$/H$\beta$, and [NeIII]$3870$/[OII]$\lambda \lambda$ $3726$-$3729$.
From these ratios, Z$_\text{gas}$ is inferred as the value that minimizes the $\chi^2$ in the following formula :
\begin{equation}\label{eq_zgas}
\chi^2(x)=\sum_{i}^{}\frac{(\mathrm{R}_{\mathrm{obs,i}}-\mathrm{R}_{\mathrm{cal,i}}(x))^2}{(\sigma_{\mathrm{obs,i}}^2+\sigma_{\mathrm{cal,i}}^2)}.
\end{equation}
where $x=\mathrm{12+log(O/H)}=\mathrm{log}(Z_{\mathrm{gas}}/\mathrm{Z}_{\odot})+8.69$, $\mathrm{R}_{\mathrm{obs,i}}$ and $\mathrm{R}_{\mathrm{cal,i}}(x)$ are the logarithm of the $i$-th observed line ratio and its predicted value at $x$ according to the calibrations of \citet{bian18}, while $\sigma_{\mathrm{obs,i}}$ and $\sigma_{\mathrm{cal,i}}$ are the uncertainty on $\mathrm{R}_{\mathrm{obs,i}}$ and the calibration uncertainty, respectively. 
The $1\sigma$ uncertainties on Z$_\text{gas}$ were estimated by perturbing the observed line ratios by their $1\sigma$ errors, recalculating Z$_\text{gas}$ $500$ times, and finally taking the 68th percentile width of all the obtained values. 
We remark that the calibrations of \citet{bian18} were specifically derived for our redshift range and that they provide results that are quantitatively consistent with values obtained from the direct method, as shown in \citet{sanders20}.

In case not all the above lines are detected, a metallicity derivation was equally performed if at least [OIII]$\lambda 5007$ \AA\ and [OII]$\lambda \lambda$ $3726$-$3729$ \AA\ were available. In particular, 6/21 galaxies have metallicities derived using all the three line ratios introduced above, 11/21 metallicities were derived using O32 and [OIII]$\lambda 5007$/H$\beta$, and 4/21 using only the O32 index. In spite of the different number of indices adopted, \citet{sanders20} found that using a lower number of line ratios than the complete set does not bias the solution on average, but only slightly reduces its S/N.
More specific details about the observations, data reduction, line measurements and metallicity calibrations of the NIRVANDELS subset can be found in \citet{cullen21}.

% Results
\section{Results}\label{results}

In this section, we study the stellar mass - metallicity relation (MZR) of star-forming galaxies at redshifts $2<z<4$ as a function of their environment. We use different definitions of environment, as presented in the previous section, and we analyze both the stellar and the gas-phase metallicity.

\subsection{Environmental dependence of the stellar-phase MZR}\label{MZRenvironment}

\begin{figure}[t!]
    \centering
    \includegraphics[angle=0,width=0.92\linewidth,trim={0.3cm 0cm 0.2cm 0.1cm},clip]{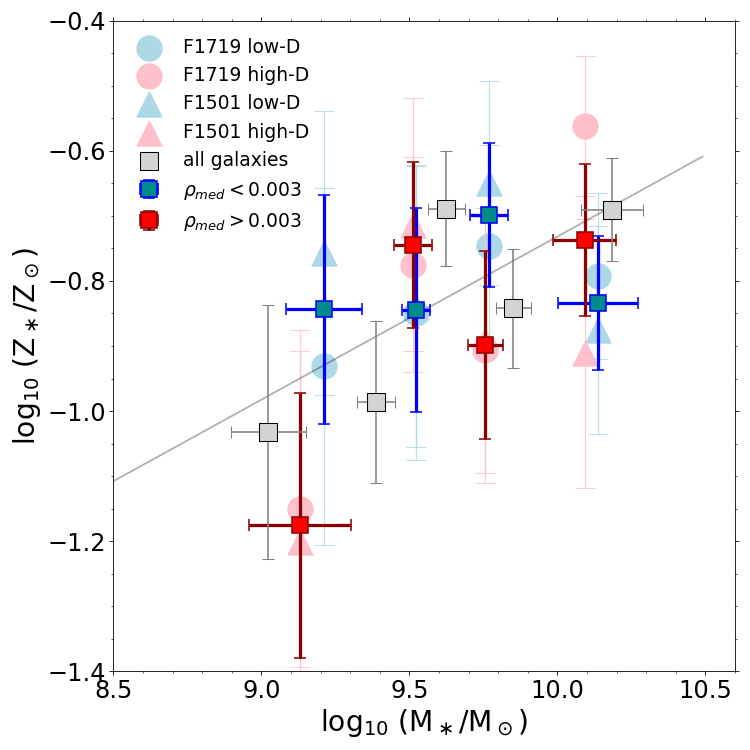}
    \includegraphics[angle=0,width=0.92\linewidth,trim={0.3cm 0cm 0.2cm 0.1cm},clip]{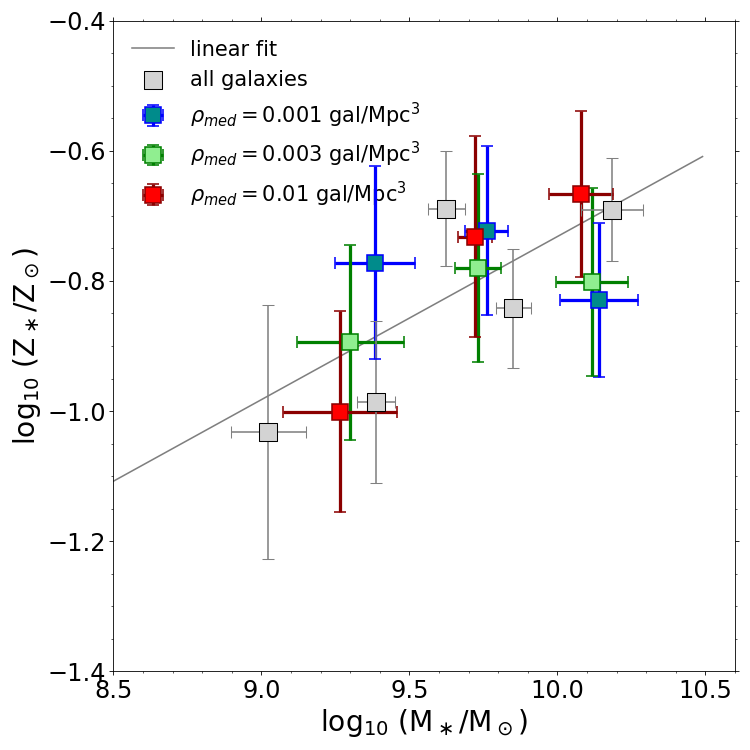}
    \vspace{-0.2cm}
    \caption{\small \textit{Upper panel:} Stellar MZR of the whole population of star-forming galaxies at $2<z<4$ (gray squares), with its best-fit relation (gray continuous line). We overplot the MZR color coded according to the local galaxy density being lower (blue squares) or higher (red squares) than the sample median of $0.003$ gal/Mpc$^3$, with the linear relations fitted to each subset (fixing the slope to the global MZR value), and drawn with the corresponding colors. The shaded area around the best-fit relations represent the $1\sigma$ uncertainty for the MZR normalization. The big square symbols represent the median stellar mass of the galaxies in each bin and the stellar metallicity derived from their spectral stack, as explained in the text. The big shaded triangles and circles indicate instead the metallicities obtained from each of the two features (respectively, at $1501$ and $1719$ \AA) that contribute to the average metallicity value in the bin. The horizontal and vertical error bars of the big squares represent, respectively, the standard deviation of M$_\star$ in each bin and the $1\sigma$ uncertainty associated to the metallicity measurement. \textit{Bottom panel:} Same as above, but using three bins of local density, with the intermediate density bin drawn in green. The legend reports the median local density for each of the three subsets. % overdensity flag (as defined in the text) in the left panel, 
    }\label{MZR_environment1}
    \vspace{-0.3cm}
\end{figure}

\begin{figure*}[t!]
    \centering
    \includegraphics[angle=0,width=0.85\linewidth,trim={0cm 1.5cm 10cm 0.1cm},clip]{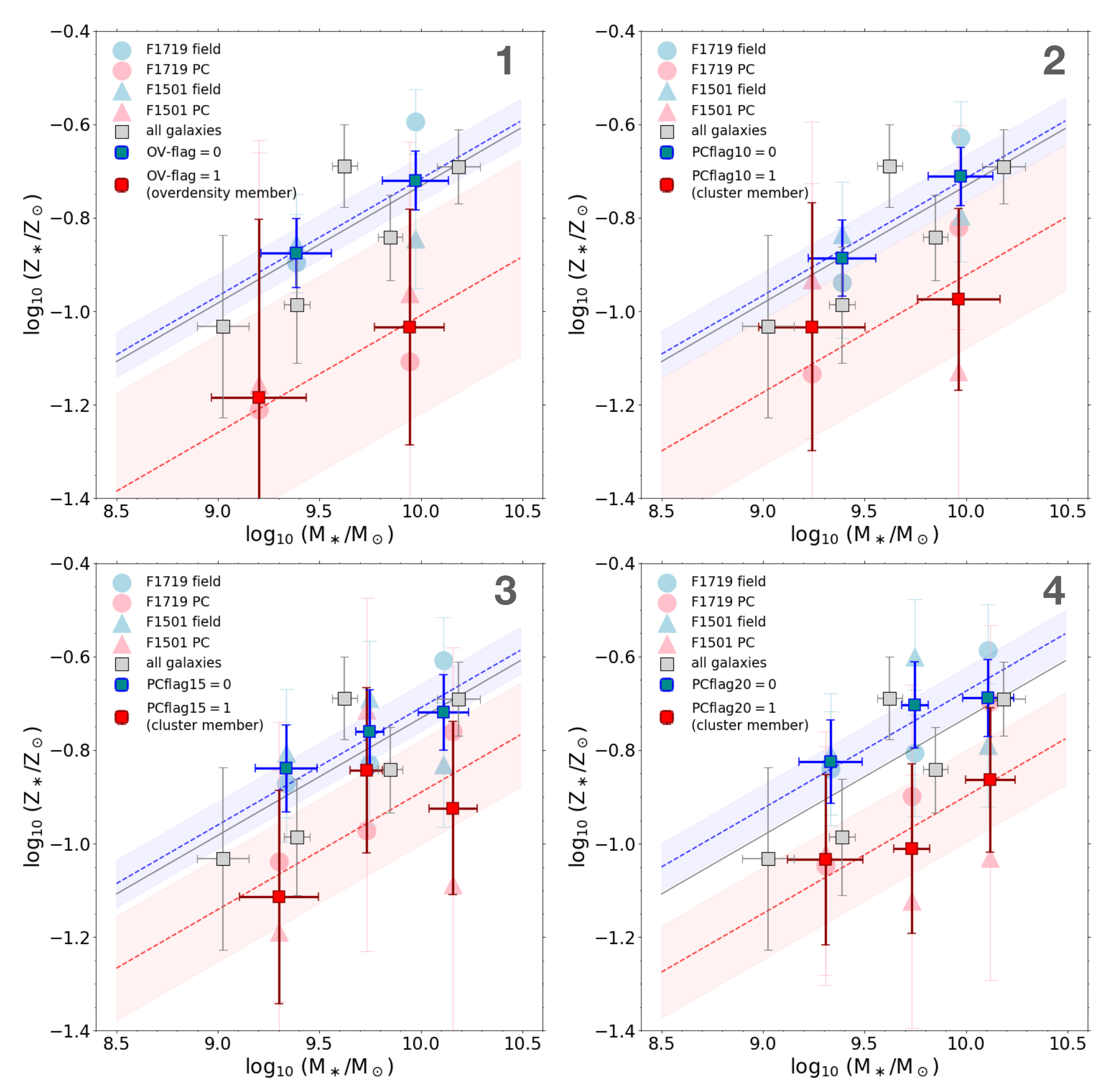}
    \caption{\small Stellar mass-metallicity relation color coded according to the overdensity flag (panel 1) and protocluster membership, as defined in the text, with increasing radii from the overdensity structure centers ($10$, $15$, and $20$ cMpc from panel 2 to 4). Gray squares, dashed line and colored filled markers are the same as in Fig. \ref{MZR_environment1}. %We overplot the MZR of the whole population of star-forming galaxies in the redshift range $2<z<4$ (gray squares), with the best-fit (dashed gray) line. 
    }\label{MZR_environment2}
\end{figure*}

\begin{figure*}[t!]
    \centering
    \includegraphics[angle=0,width=1\linewidth,trim={0cm 39cm 0cm 0cm},clip]{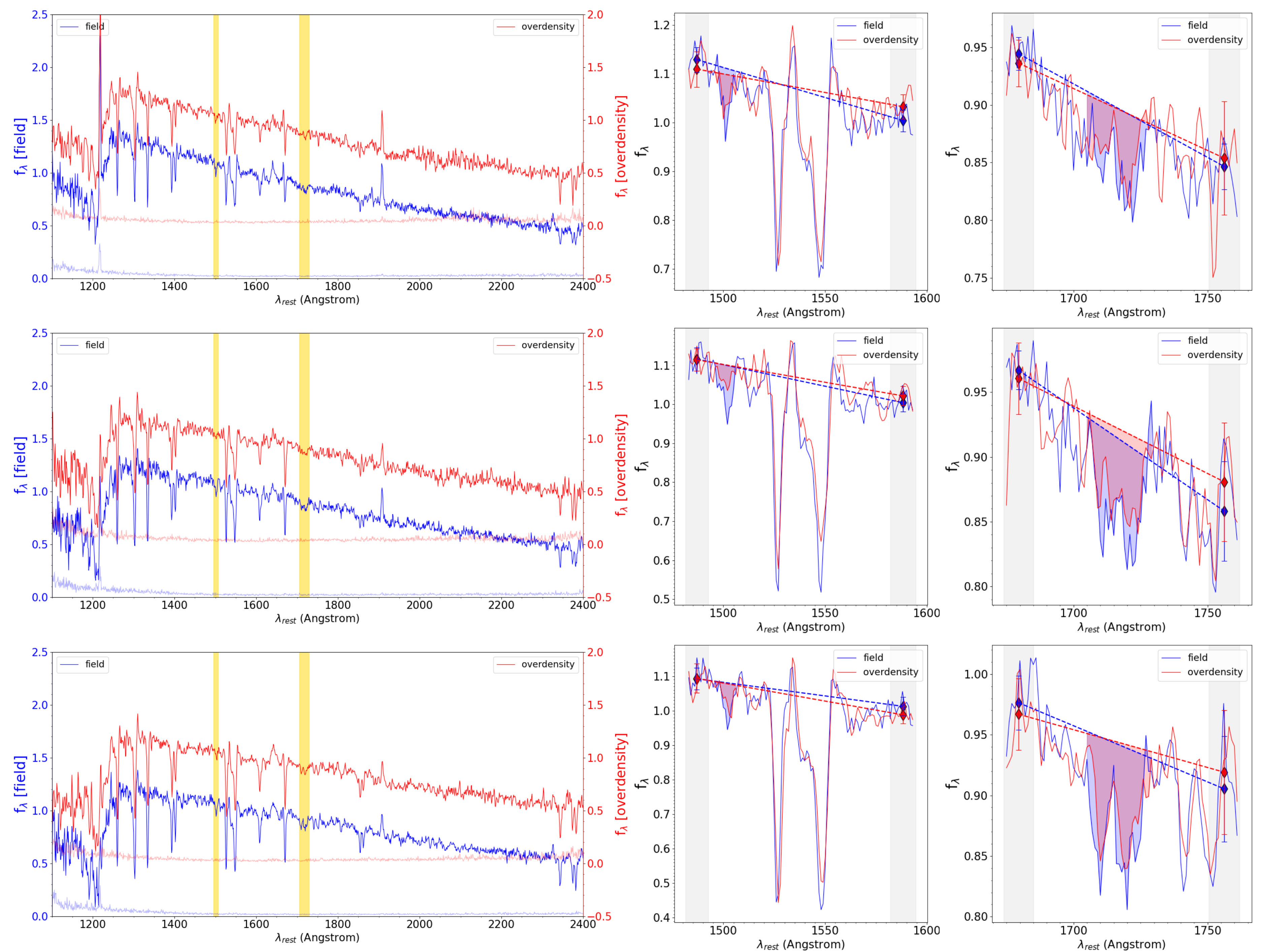}
    \caption{\small Stacked spectra corresponding to the bins in stellar mass and PC-flag$_{20}$ adopted in panel $4$ of Fig. \ref{MZR_environment2}. In each row are shown, for each stellar mass bin (increasing mass from top to bottom), the stacks for the galaxy population inside the overdensity structures (in red) and in the field (in blue). For visualization purposes, the two scales of the fluxes and corresponding errors (on the left and right hand y-axis, respectively), have been slightly shifted. The wavelength ranges corresponding to the $1501$ and $1719$ metallicity indices are highlighted in yellow in the main plot, while they are further magnified in two panels on the right of each row. Here the features corresponding to the two galaxy populations (protocluster and field) are overplotted to facilitate the comparison. The blueward and redward shaded gray regions correspond to the \citet{rix04} windows to define the pseudocontinuum across the absorption feature through a straight line fit, while the shaded red and blue areas are used for the calculation of the equivalent width. 
    }\label{Figure_spectrum}
\end{figure*}

As a first step of our analysis, we stack the spectra of galaxies in the same stellar mass bins defined in our previous work (C21) to update the global MZR, considering the last VANDELS data release (DR4). The result is shown in Fig. \ref{MZR_environment1}, where the gray squares are the metallicities calculated in each mass bin, while the gray dashed line indicates their best fit linear relation. Overall, we find an increase of $\sim0.5$ of the stellar metallicity between M$_\star =$ $10^9$ and $10^{10.5}$ M$_\odot$, as already showed in C21. The individual estimates are also fully consistent within $1\sigma$ uncertainty with the previous estimates.

%In Fig. \ref{MZR_environment1}-\textit{top panel}, 
We then divide the whole galaxy sample in two bins depending on whether their local density values $\rho_\text{gal}$ are lower or higher than the population median $\rho_\text{gal}=0.003$ gal/Mpc$^3$. For each subset of $\rho_\text{gal}$, we further divide galaxies in four bins of stellar mass, ensuring enough statistics and an equal number of objects in each stellar mass bin over the whole range $8.5<$ log$_{10}$ (M$_\star$/M$_\odot$) $<10.5$, as shown in Fig. \ref{MZR_environment1}-\textit{top panel}. 
We find that the MZR of galaxies in higher density and lower density regions are in agreement within $2\sigma$ with no evident systematic offsets. However, we notice that at the lowest stellar masses ($<10^{9.5}$ M$_\odot$), galaxies residing in overdense regions have a metallicity approximately $0.35$ dex lower compared to more isolated galaxies. % at fixed M$_\star$. 
In this low-mass regime, the metallicity estimate for overdense galaxies is $\sim 1\sigma$ below the global best-fit MZR. This environmental effect at low stellar masses is also found for the two metallicity indices separately, which agree rather well, while the individual values are more scattered at higher masses. 

We then divide the whole sample in three groups of increasing local density by using the third quartile of the density distribution, reducing consequently the number of bins along the stellar mass direction to keep enough statistics in each subset. The results of this exercise are shown in Fig. \ref{MZR_environment1}-\textit{bottom panel}, and reveal that the large difference in metallicity at low-masses (compared to both the global relation and to undersende regions) is driven by galaxies falling in the densest structures with $\rho_\text{median} \sim 0.01$ gal/Mpc$^3$, that is, more closely associated to the dense cores.  
For the blue squares in both panels of Fig. \ref{MZR_environment1}, even though a linear fit could suggest that the MZR in underdense regions has a completely different shape than the global relation (i.e., a substantially flatter slope), a $\chi^2$ analysis yields $\chi^2/\nu =0.68$  ($-0.45\sigma$), indicating that the global MZR cannot be ruled out (i.e., there is a $\sim60 \%$ chance to obtain a worse $\chi^2$ if the hypothesis is correct). This means that actually the blue square measurements are consistent with galaxies being drawn from the whole population (gray squares). 

We then repeat the same analysis considering the other density indicators described in section \ref{densityfield}, and binning the galaxies according to these proprerties. First, we color code the stellar phase metallicity versus mass diagram by the overdensity flag, that is, depending on whether the galaxies lie inside (OV-flag $=1$) or outside (OV-flag $=0$) of the overdensity structures. For each subset, we additionally create two stellar mass bins, getting in total $48$ ($463$) galaxies in the low mass regime, and $30$ ($332$) galaxies in the higher mass bin, respectively inside and outside of the overdensity structures. As shown in the first panel of Fig. \ref{MZR_environment2}, systems in overdensities have on average a $\sim0.3$ dex lower stellar metallicity at fixed M$_\star$ compared to the field. Because of the large uncertainty of the red squares, due to the poorer statistics in the densest regions, the significance of this offset is at the level of $1\sigma$. 
%Even though with a larger uncertainty due to the poorer statistics, we find here a slightly lower stellar metallicity for objects in overdense structures at fixed M$_\star$, but in this case the environmental effects are not significant, since the offset between the two MZRs is consistent with $0$ within $1\sigma$. 
In the last three panels of Fig. \ref{MZR_environment2}, we show with red squares the MZR of galaxies residing in protocluster structures defined with different radii of $10$, $15$, and $20$ cMpc radius (from panel $2$ to $4$), and compare it to the relations followed by systems outside of the  protoclusters (blue squares) and to the global population. In the first case, a PC-flag$_{10}=1$ traces the densest protocluster cores identified in the redshift range between $2$ and $4$. To mitigate the lower statistics in protoclusters, we divide the sample only in two stellar mass bins according to the median sample value (M$_{\star,median} \simeq 10^{9.7}$ M$_\odot$), which allows to derive final Z$_\star$ values with uncertainties $\leq 0.3$ dex. We have in this case $80$ and $44$ galaxies in the lower and higher stellar mass bin, respectively, while $431$ and $318$ systems are in the external regions with PC-flag$_{10}=0$. 
For radiii of $15$ and $20$ cMpc, we are considering protocluster members to larger distances compared to the previous case (radius $10$ cMpc). 
The advantage is that the statistics is higher, and we can thus divide the sample in $3$ equally populated bins in mass instead of only $2$. In the case where PC-flag$_{15}$ is considered, we have now $124$, $45$, and $48$ galaxies in protoclusters, from the lower to the higher stellar mass bin. 

\begin{table*}[t!]
\centering{\textcolor{blue}{Stellar-phase MZR}}
\renewcommand{\arraystretch}{1.5}  % # 2 if you want to increase cell height
\vspace{-0.2cm}
\begin{center} { \normalsize
\begin{tabular}{ | m{11em} | m{1.6cm} | m{1.7cm}| m{1.7cm}| m{1.7cm}| m{1.7cm}| m{2.1cm} | } 
  \hline
  %\textbf{Environment} & \multirow{2}{*}{\parbox[t]{\hsize}{\textsc{range of $\log$(M$_\star$/M$_\odot$)} }} & $\Delta$($\log$ Z$_\star$/Z$_\odot$) \\ 
  \textbf{Environmental diagnostic} & $\log$(M$_\star$/M$_\odot$) range & EW$_{1501,OV}$ & EW$_{1501,F}$ & EW$_{1719,OV}$ & EW$_{1719,F}$ & $\Delta$($\log$ Z$_\star$/Z$_\odot$) \\ 
  \hline
  %\textsc{local density $\delta$ (higher $-$ lower than $\delta_{median}$)} & cell2 & cell3 \\ 
  \multirow{4}{*}{\parbox[t]{\hsize}{\textsc{local density $\delta$} }} & $8.4$-$9.4$ & $0.29\pm0.16$ & $0.55\pm0.16$ & $0.85\pm0.28$ & $1.12\pm0.37$ & $-0.3\pm0.3$ \\
    & $9.4$-$9.65$ & $0.59\pm0.15 $ & $0.50\pm0.15 $ & $1.34\pm 0.25$ & $1.23\pm 0.3$ & $0.1\pm0.2$ \\
    & $9.65$-$9.9$ & $0.46\pm0.13 $ & $0.64\pm0.13 $ & $1.15\pm 0.28$ & $1.39\pm 0.24$ & $-0.2\pm0.2$ \\
    & $9.9$-$10.8$ & $0.45\pm0.13 $ & $0.47\pm0.10 $ & $1.7\pm 0.2$ & $1.32\pm 0.2$ & $0.1\pm0.2$ \\
  \hline
  \multirow{2}{*}{\textsc{overdensity structure}} & $8.4$-$9.7$ & $0.31\pm0.2$ & $0.48\pm0.07$ & $0.78\pm0.6$ & $1.17\pm0.14$ & $-0.3\pm0.4$ \\
    & $9.7$-$10.8$ & $0.42\pm0.19$ & $0.49\pm0.07$ & $0.9\pm0.4$ & $1.64\pm0.12$ & $-0.3\pm0.2$ \\
  \hline
  %\textsc{protocluster (r=10 cMpc) $-$ field} & cell2 & cell3 \\ 
  \multirow{2}{*}{\parbox[t]{\hsize}{\textsc{protocluster (r=10 cMpc)}}} & $8.4$-$9.7$ & $0.4\pm0.3$ & $0.50\pm0.08$ & $0.9\pm0.4$ & $1.11\pm0.16$ & $-0.15\pm0.3$ \\
    & $9.7$-$10.8$ & $0.32\pm0.16$ & $0.53\pm0.07$ & $1.3\pm0.3$ & $1.58\pm0.16$ & $-0.3\pm0.2$ \\
  \hline
  \multirow{3}{*}{\parbox[t]{\hsize}{\textsc{protocluster (r=15 cMpc)}}} & $8.4$-$9.6$ & $0.29\pm0.17$ & $0.52\pm0.10$ & $0.98\pm0.37$ & $1.2\pm0.18$ & $-0.3\pm0.25$ \\
    & $9.6$-$9.9$ & $0.59\pm 0.18$ & $0.61\pm 0.10$  & $ 1.07\pm0.3$ & $1.26\pm0.19$ & $-0.1\pm0.2$ \\
    & $9.9$-$10.8$ & $0.34\pm 0.17$ & $0.5 \pm 0.10$ &  $1.36\pm0.28$ & $1.62\pm0.16$ & $-0.2\pm0.2$ \\
  \hline
  \multirow{3}{*}{\parbox[t]{\hsize}{\textsc{protocluster (r=20 cMpc)}}} & $8.4$-$9.6$ & $0.38\pm0.15$ & $0.52\pm0.1$ & $0.97\pm0.3$ & $1.25\pm0.18$ & $-0.2\pm0.25$ \\
    & $9.6$-$9.9$ & $0.33\pm0.14$ & $0.68\pm0.10$ & $1.16\pm0.3$ & $1.30\pm0.2$ & $-0.3\pm0.3$ \\
    & $9.9$-$10.8$ & $0.38\pm0.15$ & $0.53\pm0.10$ & $1.47\pm0.3$ & $1.66\pm0.18$ & $-0.2\pm0.2$ \\ 
  \hline
\end{tabular} }
\end{center}

\caption{\small Table indicating, for the stellar MZR and for different stellar mass bins and environmental diagnostics, the equivalent width of the $1501$ and $1719$ metallicity indices in the spectral stacks corresponding to the overdense regions (EW$_{1501,OV}$ and EW$_{1719,OV}$) and in the field (EW$_{1501,F}$ and EW$_{1501,F}$). All EWs are in Angstrom and we assume the convention of positive values in absorption. The last column table indicates the difference in metallicity $\Delta$($\log$ Z$_\star$/Z$_\odot$) between galaxies in overdensities (or protoclusters) and in the field in each stellar mass bin.}
\label{tabella1}
\end{table*}

Considering all the four panels of Fig. \ref{MZR_environment2}, we can see that the MZR of galaxies in the field is remarkably consistent with the best linear fit derived for the whole population. In contrast, the stellar metallicities of protocluster or overdensity members are on average lower by $0.1$-$0.3$ dex, and the offset does not depend on the stellar mass range of the galaxies. 
The difference between the stellar metallicities of the two subsets in all the stellar mass bins is significant at $1\sigma$ level on average. 

Furthermore, in order to evaluate the overall significance of the offset in a different quantitative way, we fit the blue and red squares with a first-order polynomial, fixing the slope to the value derived for the global population (gray continuous line). We then compare the best-fit absolute metallicity normalizations among the different subsets. We find that the difference of the MZR normalizations between protocluster galaxies and in the field is marginally significant at $ \geq 1\sigma$ level in all the cases, and $\sim 2\sigma$ for PC-flag$_{20}$. 
We remark that the errors on Z$_\star$ and our results do not vary significantly when modifying the mass bin separation limits by $<0.1$ dex, which corresponds to the typical uncertainties on M$_\star$ from SED fitting, and they are robust also against variations of the stacking methodology, for example, on whether a $3$ or $5$ sigma clipping is performed prior to the median flux estimation at each wavelength, as in \cite{cullen19}.

Furthermore, in Figures \ref{MZR_environment1} and \ref{MZR_environment2} we also show with shaded colored markers the metallicity values obtained for each individual index (triangles for the $1501$ feature and circles for the $1719$ index), which contribute to the average value in each bin (colored squares). We notice that the estimates from the two indices are overall consistent within their $1\sigma$ uncertainties. Given the larger errors compared to the average values, they are also more scattered around the best-fit linear relations. However, even when the individual indices are considered separately, they suggest a slightly lower stellar metallicity for galaxies inside overdense regions, with the offset being more robust in the lowest stellar mass bin.

In Fig. \ref{Figure_spectrum}, we compare the stacked spectra and the metallicity indices between the galaxy population in protoclusters with radii of $20$ cMpc, where we have more statistics, and in the field (PC-flag$_{20}=0$), for all the three stellar mass bins defined above. In this example we can visually appreciate from a single measurement that the absorption features of galaxies in protoclusters (red line) are in general slightly less deep compared to those in the field (blue line), which is indicative of a lower stellar metallicity, confirming both the data displayed in Fig. \ref{MZR_environment2}- panel $4$ and the low significance of this environmental effect given the available S/N. In Fig.~9 of our previous paper C21, we have shown that the different level of absorption of the two metallicity indices is much more evident on a visual inspection when we consider a wider range of metallicities than the offset we obtain here.

Overall, we find that the stellar metallicity does not depend on the local galaxy density for the stellar mass range $9.3 < \log (M_\star/M_\odot) < 10.5$. On the other hand, galaxies at lower stellar masses show a lower stellar metallicity in overdense regions at $1\sigma$ significance. %, yielding a MZR that is substantially steeper (at $1\sigma$ significance) than in underdense regions. 
Similarly, we find a $\sim 1 \sigma$ evidence suggesting that galaxies inside protocluster structures defined around the densest cores are more metal poor compared to equally massive systems in the field, regardless of the protocluster size considered. 
In table \ref{tabella1} we summarize our results for the stellar-phase mass-metallicity relation, showing for each stellar mass bin the EW of the $1501$ and $1719$ indices in overdense and underdense regions, and, in the last column, the metallicity offset between galaxies in different environments.

\subsection{Environmental dependence of the gas-phase MZR}\label{gasphase_MZR}

\begin{figure}[t!]
    \centering
    \includegraphics[angle=0,width=0.9\linewidth,trim={7cm 5cm 17cm 2cm},clip]{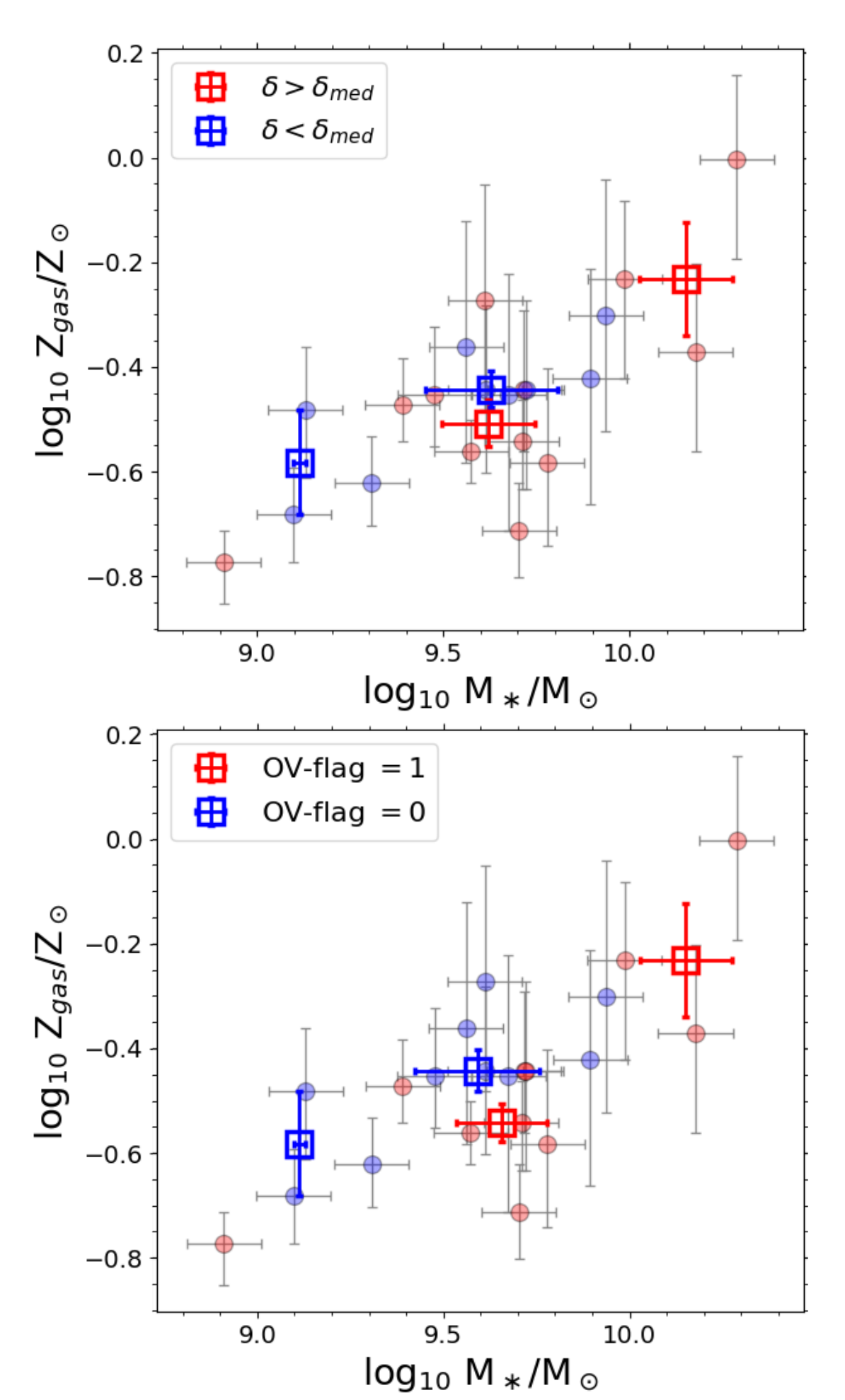}
    \caption{\small Gas-phase mass-metallicity relation for $21$ star-forming galaxies at redshifts $3<z<3.8$ from the NIRVANDELS survey, color coded according to galaxy density (higher or lower than the sample median $=0.003$ gal/Mpc$^3$) and to the overdensity flag, in the upper and lower panel, respectively. The bigger colored squares represent the median gas-phase metallicities and the median stellar mass of the galaxies in the same bin. The vertical and horizontal error bars represent, respectively, the $1\sigma $ uncertainty of the median metallicity and the standard deviation of stellar masses in each bin. We do not display a median point in case where a bin is composed by only one galaxy. % and protocluster membership (assuming a radius of $10$,$15$, and $20$ cMpc and $\Delta z=0.04$). 
    }\label{gasMZR_environment_figure}
    \vspace{-3mm} 
\end{figure}

\begin{table}[t!]
\centering{\textcolor{blue}{Gas-phase MZR}}
\vspace{-0.2cm}
\begin{center} { %\small
\begin{tabular}{ | m{11em} | m{1.6cm}| m{2.2cm} | } 
  \hline
  %\textbf{Environment} & <$\log$ (M$_\star$/M$_\odot$)> & $\Delta$($\log$ Z$_{gas}$/Z$_\odot$) \\ 
  \textbf{Environmental diagnostic} & range of $\log$(M$_\star$/M$_\odot$) & $\Delta$($\log$ Z$_{gas}$/Z$_\odot$) \\ 
  \hline
  \textsc{local density $\delta$ } & $9.3$-$9.9$ & $-0.07 \pm 0.06$ \\ 
  \hline
  overdensity structures & $9.3$-$9.9$ & $-0.10 \pm 0.05$ \\ 
  \hline
\end{tabular} }
\end{center}

\caption{\small Gas-phase MZR as a function of local galaxy density (higher or lower than the median sample) and overdensity flag. We highlight in the third column the difference in metallicity $\Delta$($\log$ Z$_{gas}$/Z$_\odot$) between galaxies in overdensities and in the field in each stellar mass bin, for which we give the corresponding range in the second column.}
\label{tabella2}
\end{table}

We investigate now the environmental effects on the metal content of the ionized gas. To this aim, we derive the relation between the stellar mass and gas phase metallicity for the subset of $21$ star-forming galaxies at redshifts $2.95 < z < 3.8$ that are taken from the NIRVANDELS follow-up, as presented in Section \ref{gas-metallicity}. As done for the stellar metallicity, we divide this sample in two bins according to their local galaxy density $\rho_{gal}$ and overdensity flag (OV-flag). In each bin, we then calculate Z$_\text{gas}$ as the median gas-phase metallicity of all the galaxies residing in that bin, while the uncertainty on the median is derived as its standard error ($=1.25\times \sigma/ \sqrt[2]{N_{bin}}$), where $\sigma$ is the standard deviation of the metallicity values in the sample and N is the number of objects in the bin.

The results are shown in Fig. \ref{gasMZR_environment_figure}, where we compare the MZR of galaxies residing in high density regions ($\rho_\text{gal}$ above the sample median of $0.003$ gal/Mpc$^3$) or inside the identified overdensity structures (OV-flag $=1$), to that of more isolated systems. 
In the central stellar mass bin from $10^{9.4}$ to $10^{9.9}$ M$_\odot$, where we have more statistics, we see that galaxies in overdense regions have a $\sim 0.1$ dex lower metallicity Z$_\text{gas}$ than in the field, with a significance of at least $1\sigma$ in the first case and $2\sigma$ in the second panel. We do not extend the median relation to the lowest and highest stellar mass regimes, respectively for the subset inside and outside overdensities, because the median metallicity value would be based on just one object.  
In any case, we do not notice a significant variation of the slope of the MZR for the different subsets. We do not analyze the case with the protocluster flag classification because of the very low number statistics available in each bin.
Overall, we remark that, despite the lower statistics available to test the gas-phase MZR compared to the stellar-phase analog, we see a significant metallicity difference for the majority of our galaxies (i.e., around the median stellar mass of the sample of $10^{9.7}$ M$_\odot$) as a function of the local density, with more metal poor systems residing preferentially in overdense regions and structures identified in Section \ref{densityfield}.
In table \ref{tabella2} we show our results for the gas-phase mass-metallicity relation, specifying the metallicity difference between galaxies in overdense and underdense regions in the middle stellar mass bin.

%%%%%%%%%%%%%%%%%%%%%%%%%%%%%%%%%%%%%%%%%%%%%%%%%%%%
%%%%%%%%%%%%%%%%%%%%%%%%%%%%%%%%%%%%%%%%%%%%%%%%%%%%
%%%%%%%%%%%%%%%%%%%%%%%%%%%%%%%%%%%%%%%%%%%%%%%%%%%%

\section{Discussion}\label{discussion}

In this section, we compare our results to other recent findings in the literature at similar redshifts. We then investigate the origin of the weak environmental dependence of the mass-metallicity relation and give possible physical interpretations for this. With this aim, we better characterize the population of galaxies in overdense and underdense regions by looking at other critical parameters derived through SED fitting or from our previous VANDELS paper C21. 
Finally, we compare our findings to the predictions of different semi-analytic models (SAM), following the approach that we already adopted in C21, and to the outcomes of two well known hydrodynamical simulations. We consider different methods to characterize the environment, discussing how it affects the mass-metallicity relation $2$ billion years after the Big Bang, and what are the limitations of the models. 

\subsection{Physical interpretation of the metallicity offset}\label{discussion1}

\begin{figure}[ht!]
    \centering
    \includegraphics[angle=0,width=0.99\linewidth,trim={0.cm 0.cm 0.cm 0.cm},clip]{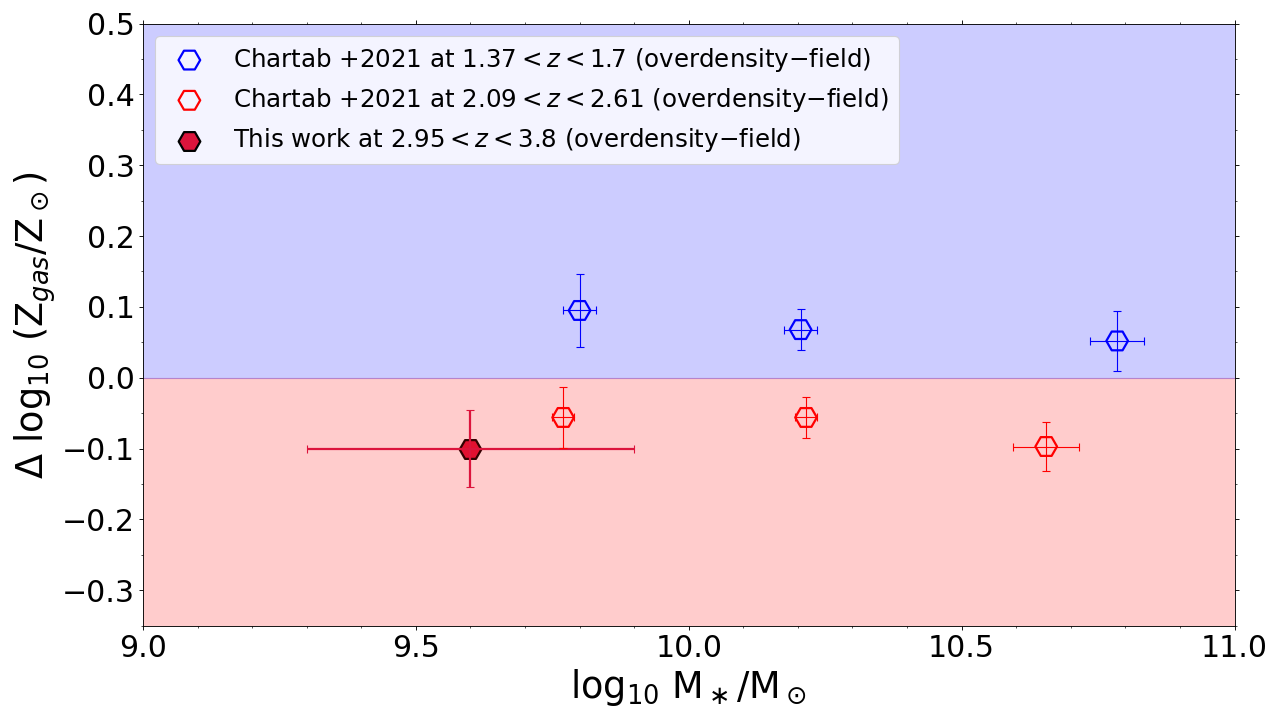} 
    \caption{\small Diagram comparing the gas-phase metallicity offset between overdense and underdense regions in the stellar mass range between $10^8$ and $10^{11}$ M$_\odot$. Our data is shown with a red filled symbol, overplotted to recent findings from the MOSDEF survey at redshifts below and above $2$.}\label{MZR_gas_comparison_other_works}
\end{figure}

\begin{figure*}[ht!]
    \centering
    \includegraphics[angle=0,width=0.92\linewidth,trim={0.cm 0.cm 0.cm 0.04cm},clip]{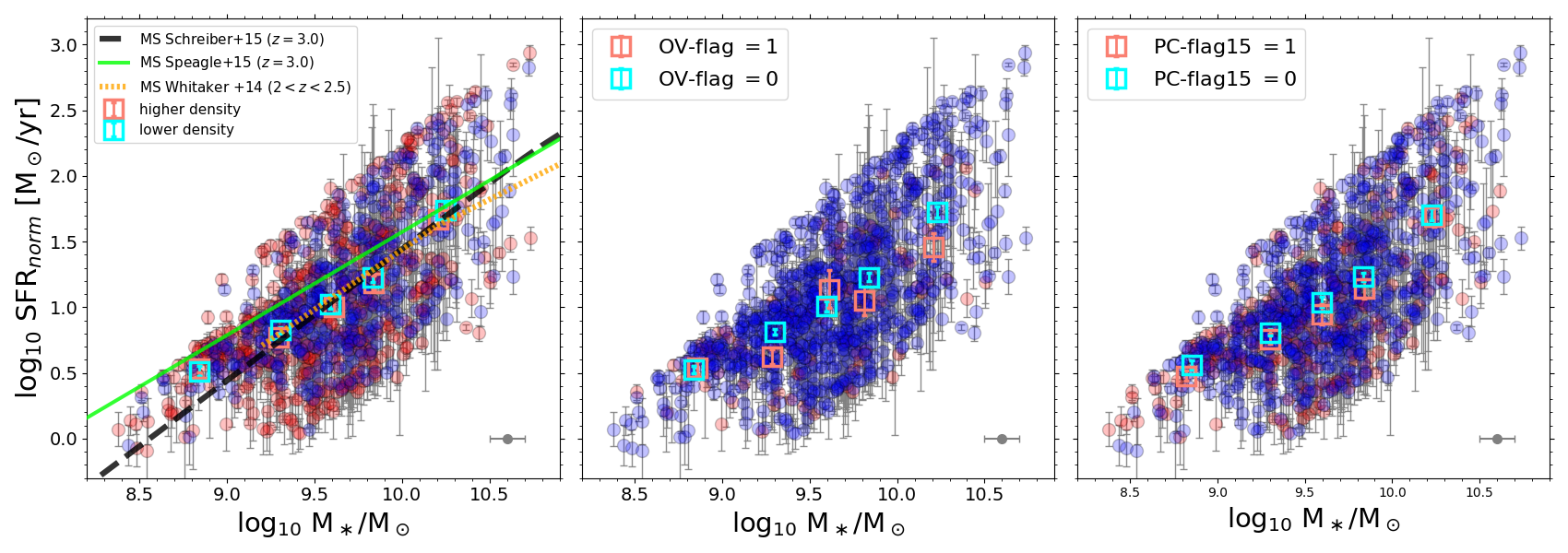}
    \includegraphics[angle=0,width=0.93\linewidth,trim={0.cm 0.cm 0.cm 0.4cm},clip]{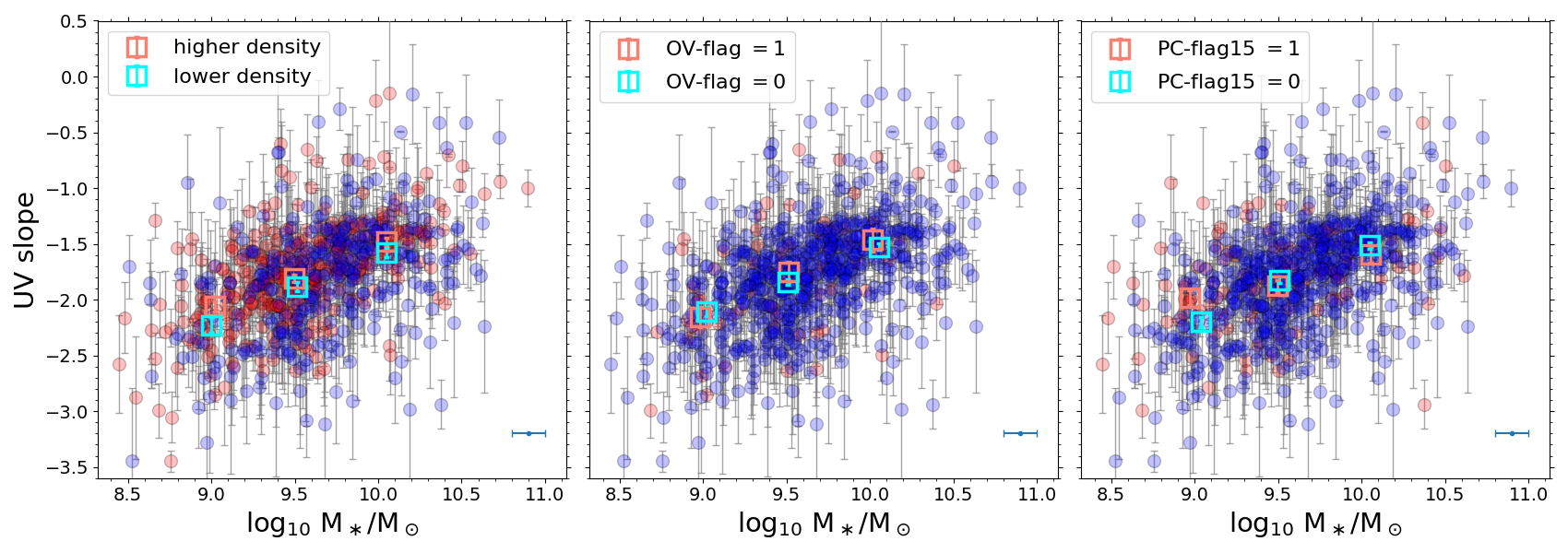}
    \caption{\small Environmental dependence of the SFR - stellar mass and $\beta$ - mass relations for the VANDELS galaxies analyzed in this work, respectively in the first and second row. The colors of the symbols in each plot are based, from left to right, on the local density (higher or lower than the sample median), overdensity flag, and protocluster membership PC-flag ($15$ cMpc). The Main Sequence of star-formation at redshift $\sim 3$ from different works is also shown for comparison in the first plot. The SFR has been normalized to $z=3$ following the $\sim (1+z)^{2.8}$ redshift dependence of \citet{sargent14}. The typical uncertainties on the stellar masses from SED fitting are shown in the bottom right corner of each plot.}\label{comparison_other_parameters}
\end{figure*}

We have seen in previous sections that the mass-metallicity relations, both the stellar and the gas-phase versions, have a weak dependence on the environment, reaching a significance of $\sim 2\sigma$, according to which galaxies in overdensities and protocluster structures tend to be less metal rich compared to equally massive systems in the field. 
It is also true that the systematic metallicity offset between galaxies inside and outside of the overdensities (Fig. \ref{MZR_environment2}) is not found in the whole stellar mass range as a function of the local galaxy density (Fig. \ref{MZR_environment1}), suggesting that the environmental effects should be more important at lower masses in the proximity of densest protocluster cores. 

Regarding the stellar-phase MZR, this work represents the first systematic investigation of environmental effects on this relation at high redshift, and it suggests that there might be a mild evolution compared to the local universe, with the environment playing some role at $z\gtrsim 2$, although this result is still low S/N in nature. However, we remind that the analysis in the local universe, which shows no environmental dependence at all as a function of density and clustercentric distance for the star-forming galaxy population \citep{trussler21}, is based on optical absorption lines, which trace more evolved stellar populations, hence not directly comparable to our work. 
As far as the gas-phase MZR is concerned, our results are in agreement with the environmental trend reported by \citet{chartab21} from the MOSDEF survey, suggesting that the downward metallicity offset of $\sim0.1$ dex that they found on average between $\sim10^{9.7}$ and $10^{10.7}$ M$_\odot$ can be extended to slightly lower masses, down to $\sim10^{9.3}$ M$_\odot$, as shown in Fig. \ref{MZR_gas_comparison_other_works}. 
It is also not very surprising that a similar environmental dependence, with less metal rich galaxies in overdensities, is found simultaneously for the stellar and gas-phase MZR. Indeed, our metallicity indicators trace the metal content of very young, massive O-B stars, thus more closely associated to the surrounding gas from which they formed. 

In spite of its low S/N, it is useful to understand where this metallicity offset ($\sim 0.1$-$0.3$ dex) can possibly arise. We first look for potential biases in our metallicity estimations. As shown in C21, the S/N of the stellar continuum strongly affects the uncertainty on the EW measurements (hence on the metallicity error). However, using the same simulations, as described in the Appendix \ref{appendix1}, we find that the S/N does not bias the metallicity value itself (Fig. \ref{Zvalue_SNR}).

After excluding systematic measurement effects, we now characterize other physical properties of our galaxies in overdense regions compared to the field, in order to check whether the small offset is driven by selection effects. 
For example, the mass-metallicity relation is known to vary as a function of the SFR of the galaxies, and this relation is in place up to at least $z\sim2.5$ \citep{cresci19}. In addition, as shown in C21, the stellar metallicity correlates with the UV rest-frame luminosity (M$_{1600 \AA}$) and with the slope of the UV continuum ($\beta$), a proxy for the dust attenuation, with more metal poor galaxies being on average fainter in UV, less attenuated (i.e., with lower $\beta$), hence on average younger, because of shorter time available for chemical enrichment. Therefore, it is worth to check whether the lower stellar metallicities inside protoclusters are driven by a higher level of SFR or by a lower UV luminosity, dust attenuation, and stellar age of the galaxies selected in these structures compared to the field. 

To this aim, we explore the environmental effects on different relations involving the stellar mass M$_\star$ and other physical parameters, namely the SFR, the UV slope ($\beta$), the absolute magnitude at $1600$ \AA\ rest-frame (M$_{1600}$), and the luminosity weighted stellar age (age$_{LW}$, which is the one more directly comparable to our UV-based stellar metallicities). The UV slopes and M$_{1600}$ are inferred from the available photometry as described in C21, while the SFRs and the age$_{LW}$ are derived with Beagle through the same SED fitting procedure used for the estimation of the stellar masses in Section \ref{parent}.
We divide our sample in two groups of galaxies according to their local density (lower or higher than the sample median), overdensity flag, and protocluster membership. For the latter, we choose as representative the case with a protocluster radius of $15$ cMpc, which is a compromise between $10$ and $20$ cMpc that were also considered in the previous section. We further divide these subsets in three bins of stellar mass, which ensures a sufficient statistics for our analysis. 
In Fig. \ref{comparison_other_parameters}, it can be seen that the SFR and the UV slopes are not significantly affected by the environment at the redshift considered in this work. Galaxies in overdense regions show indeed similar levels of star-formation activity and $\beta$ (hence UV-based dust attenuation) compared to more isolated systems. Their mass-SFR and mass-$\beta$ relations are perfectly consistent within the errors, without systematic offsets. 
Similarly, we have also checked that the stellar ages and M$_{1600}$ of galaxies inside and outside of overdense structures are perfectly consistent within $1 \sigma$, suggesting for them no environmental dependence at all.
We conclude that the lower metal content of protocluster galaxies is not driven by selecting systems in overdense structures with a higher level of SF activity, a lower UV luminosity and dust attenuation, or different stellar age compared to the field, but it has likely an environmental origin. 

The environment can play a role in different ways. A higher efficiency of gas stripping phenomena and harassment in protoclusters might deprive galaxies of part of their metal-enriched gas content. 
% as a consequence of pristine gas inflows from their surroundings \citep{keres05,dekel06,ocvirk08}
In addition, major mergers and interactions, which are expected to be more frequent in dense environments \citep[e.g.,][]{gottlober01,park09,fakhouri09}, have been shown to produce strong outflows of metal enriched gas toward the CGM \citep{hani18}, hence they could also be responsible for a decrease in metallicity in the overdensities. These events can also efficiently mix their metal poor, cold gas reservoirs with metals coming from new generated stars, as found observationally at low redshift \citep{ellison08}, and predicted from cosmological simulations \citep{torrey12,bustamante18}. At high redshifts, mergers might produce only a very mild enhancement of star formation at $z \sim 3$ \citep{fensch17}, hence we are able to explain an overall decrease of metallicity of the systems without necessarily requiring a significant increase of the global SFR, which indeed is not observed in our data.
A full exploitation of HST images, and JWST in the future, will also allow a proper morphological analysis to understand the importance of mergers and interactions in overdense structures. 

Another possibility could be related to an enhanced AGN activity inside the protoclusters and overdense regions, which can induce powerful outflows ejecting enriched material to the external regions, as observed in local galaxy clusters \citep{kirkpatrick11}. However, the fraction and the properties of AGNs in different environments are mostly unknown at redshifts $>2$, and will be addressed in a following paper.
Moreover, given that the metallicity is also known to anti-correlate with the equivalent width of the Ly$\alpha$ emission line, we note that our downward metallicity offset cannot be driven by a larger fraction of Ly$\alpha$ emitters in protoclusters, because exactly the opposite was found in G20, with these systems lying preferentially outside of overdense regions. 

%\subsection{Comparison to other works}\label{comparison_section}

Finally, as mentioned above, our results are in agreement with the environmental dependence found in the analysis of \citet{chartab21} for the gas-phase MZR obtained from $\sim300$ star-forming galaxies in COSMOS, but their proposed explanation, invoking more efficient gas inflows inside overdensities, deserves a more extensive discussion. 
In their work, the scenario of cold gas inflows is supported by the evidence of a slight increase of SFR in overdensity members, even though they claim it is not significant due to the large uncertainties of their SFRs. However, another study by \citet{koyama13}, comparing a sample of $\sim50$ cluster galaxies at $z\simeq2.2$ to field galaxies from the HiZELS survey, found that the M$_\star$-SFR relation does not vary with the environment at that cosmic epoch, which is perfectly in agreement with our results at redshifts $2<z<4$. Overall, it seems that there are no striking evidences confirming observationally that cold gas inflows play the dominant role in decreasing the metallicity in overdense regions for the normal star-forming galaxy population at high redshift. However, they might still play an indirect role by gradually replenishing the cold gas reservoirs of protocluster galaxies, diluting the metal content of the gas that fuels the ongoing star-formation activity, but without significantly increasing its consumption rate (i.e., the SFR) compared to equally massive systems in the field.

In order to see a clear environmental effect on the SFR in protoclusters as due to cold gas inflows from the cosmic web, we might target a different population of galaxies with more extreme properties.
For example, highly star-forming (i.e., off-Main Sequence) systems, which can host a significant fraction of optically obscured star-formation \citep{calabro18,calabro19}, are automatically excluded from our analysis because of their faint, low S/N, UV continuum, but would be worth to study at longer wavelengths. In general, it is expected from several works to find a large excess of dust-obscured galaxies in protoclusters at redshifts above $2$ that only appear in far-IR, sub-mm and radio surveys \citep[e.g.,][]{daddi09,dannerbauer14,umehata15,daddi17}. In a recent study by \citet{daddi21}, where they observe direct evidence of filamentary, metal-poor cold gas streaming toward the center of a galaxy group at $z\sim2.91$, most of the induced star-formation activity ($1200$ M$_\odot$/yr in total) is spread among IR luminous sources, which emit a negligible fraction of their light in the UV-optical rest-frame. H-dropout galaxies, which are extreme starbursts completely missed in optical, near-infrared, and UV surveys, are also shown to be highly clustered in the early universe \citep{wang19}. %, and should be targeted with instruments at longer wavelengths.  

In conclusion, pristine gas accretion from the cosmic web into protoclusters might trigger preferentially optically obscured, starburst activity. However, it might also play a role, in combination with the other mechanisms, in the dilution of metals of normal star-forming galaxies in overdensities, without producing a sudden enhancement of their star-formation activity in the stellar mass range explored by VANDELS ($10^9< M_\star/M_\odot <10^{10.5}$). 
% ***** while we think it is not the dominant mechanism able to explain the decrease of metallicity in normal 'main-sequence' star-forming protocluster galaxies 
Probing additional physical properties of our galaxies, such as their dust and cold gas content, and looking for optically obscured star-forming components with far-infrared observations, will centainly help to understand the processes that are playing the most important role in the starting phases of galaxy clusters evolution. 
%Osservazioni future possono anche dirci quale di questi meccanismi discussi sopra e' piu' efficiente.

\subsection{Comparison of the results to theoretical predictions}\label{theoretical}

Now it is interesting to compare our findings in Section \ref{results} to the theoretical predictions of semi-analytic models (SAM) and simulations. In C21, we have compared our VANDELS star-forming galaxies at $z\sim3$ to mock light cones based on the GAlaxy Evolution and Assembly model (GAEA). %, finding an agreement between the observed and the theoretical slope of the mass-metallicity relation. 
In this section, we use the same light cones adopted in the previous paper to study the environmental effects on the MZR. Then, we extend the analysis to the full GAEA snapshot, in order to have a larger galaxy sample (hence a higher statistics), and to be able to compare on the same ground with other models, both SAMs and hydrodynamical simulations. 

In all our models and simulations, the stellar (gas) metallicity is computed as the mass fraction of all metals (i.e., elements heavier than helium) in the stellar (cold gas) component of the galaxies, hence it is mass-weighted. This weighting scheme might lead to a small offset of $\sim0.1$-$0.2$ dex compared to the observed stellar metallicity, which is luminosity weighted, as discussed in detail in \citet{cullen19}, C21, and \citet{fontanot21}. Although models and simulations reproduce well the slope of the observed MZR, this effect, in addition to generally different chemical enrichment prescriptions, explain why they are not currently able to reproduce the exact normalization of the observed MZR. However, we remark that the scope of our analysis is not to investigate the origin of these systematic discrepancies, but rather to analyze, for each type of model, the MZR offset arising only from variations of the environment.

\subsubsection{Comparison to VANDELS light cones}\label{lightcone}

\begin{figure*}[ht!]
    \centering
    \includegraphics[angle=0,width=0.95\linewidth,trim={0.cm 14.cm 0.cm 0cm},clip]{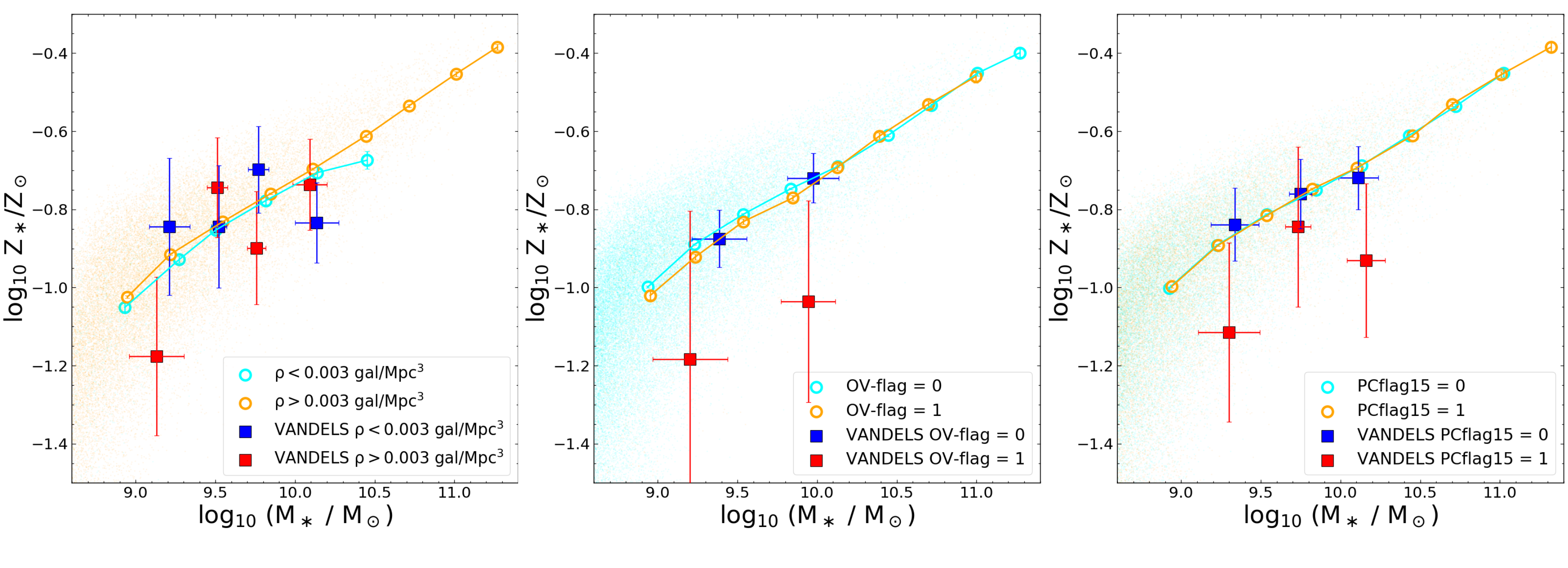}
    \caption{\small Stellar mass - stellar metallicity relation of mock galaxies from VANDELS cones, performed to reproduce the same sky area and redshift range targeted by our survey. The whole galaxy population is color coded in two bins of local galaxy density, similarly to Fig. \ref{MZR_environment1}-\textit{top}, and according to their overdensity flag and protocluster membership (second and third panel), assuming a radius of $15$ cMpc. The theoretical MZR have been offsetted by $-0.4$ dex to match the observed, global MZR normalization. The vertical error bars of the models represent the error on the median metallicity estimation in each bin.}\label{comparison_mock}
\end{figure*}

Mock galaxies light-cones have been generated inside our collaboration to mimick as much as possible our observations, in order to have the same redshift range and a similar depth and selection function to that of VANDELS, as explained in C21. While we will describe the physical details of the entire GAEA model in a following section, we want to test here the effects of the environment on the mock galaxy population considered in our previous work. For this specific subset, G20 derived the galaxy density map with a similar procedure to that described in Section \ref{densityfield}, allowing to identify mock overdensity structures and protoclusters with the same methodology already used for the observations.
We are thus able to plot the MZR for mock galaxies as a function of local density, overdensity flag and protocluster membership, with the latter defined assuming a $15$ cMpc radius from the cluster centers. 

The results are shown in the three panels of Fig. \ref{comparison_mock}.
It is immediately clear that no systematic offsets in metallicity are found, regardless of the classification adopted to distinguish between galaxies in more and less dense regions. % A possible reason for the discrepancy with the previous result on central galaxies only could be the small dynamic 
Only in the middle panel a small systematic offset of $\sim - 0.05$ dex in metallicity is present for galaxies with M$_\star < $ M$_\odot$ inside the reconstructed overdensities, even though it is not significant and by far less than the effect seen with our observations.

\subsubsection{Comparison to full suite semi-analytic models}\label{SAM}

We now investigate the environmental effects for a larger sample of galaxies taking into account the full snapshots of the models. In order to explore a wider range of model parameters and theoretical prescriptions, we also consider the SAM of \citet[][hereafter DLB07]{delucia07}. 
Since this represents the precursor of GAEA, it offers an interesting possibility to analyze different feedback scenarios than GAEA, which could have an important impact on the gas and metal circulation in the galactic environment, while keeping the same underlying cosmological framework. We introduce here the main common features of the two models. Then, we explore their peculiarities and predictions in more detail, beginning from DLB07, which was developed earlier, and we follow the various modifications that have been apported to it in chronological order. 

Both GAEA and the model of DLB07 are constructed from the dark matter halo merger trees of the Millennium Simulation \citep{springel05b}, which adopts a $\Lambda$CDM cosmology with $\Omega_\Lambda$ $= 0.75$, $\Omega_\text{m} = 0.25$, $\Omega_\text{b} = 0.045$, n $= 1$, $\sigma_8 = 0.9$, and H$_0 = 73$ km/s/Mpc, based on the WMAP1 results. As described in detail in \citet{springel05b}, they identify dark matter haloes with a standard friends-of-friends (FoF) algorithm with a linking length of $0.2$ times the main particle separation at each redshift. 
For each FoF halo, a virial mass of the structure is computed as the dark-matter mass contained inside a sphere centered in the minimum of its gravitational potential well, and with a radius $r$ such that the inner halo has an overdensity $200$ times the critical density of the Universe $\rho_\text{crit}=3$ H$_0^2$/($8\pi$G), which corresponds approximately to that expected for virialized groups \citep{croton06}.
This procedure allows to distinguish galaxies according to the virial mass of their host FoF group, that we take here as the best representative parameter to characterize the different environments quantitatively. Indeed, more massive FoF groups would likely play the role of attractors around which bigger structures will form at later epochs. The total mass of the overdensity structures identified in VANDELS ranges between $10^{11}$ and $10^{13}$ M$_\odot$ (G20), thus we will consider groups within this mass range in the following analysis. Moreover, we do not consider the very low mass halo regime (M$_\text{halo} < 10^{11}$), where we are approaching the resolution limits of the Millennium Simulation and the results might be less meaningful, especially for the satellite population. 
As we have mentioned in the introduction (Sect. \ref{introduction}), a useful distinction made in the local Universe is between central and satellite galaxies. The first are defined as the most massive galaxies in a FoF group, while all the remaining systems are dubbed satellites. They are found to have in general different properties, with the former less metal-rich than equally massive satellites \citep{pasquali10}. While we are not able to observationally discriminate between the two galaxy types in the identified VANDELS overdensity structures, it is interesting from a theoretical point of view to check the effects of the environment for these separated subsets. %, which are thus included in our analysis. 

The baryonic component is treated according to a set of physically or observationally motivated analytical prescriptions that describe the various processes involving all the baryonic phases (i.e., stars, cold and hot gas), such as gas condensation through cooling flows, star formation, stellar feedback, metal evolution, black hole growth, and AGN feedback. 
The adopted feedback scheme is one of the most distinguishing features among different SAMs and it is expected to have a strong impact on the MZR relation \citep[e.g.,][]{delucia17}. In addition it is one of the least understood process from the observational point of view, representing the ensemble of phenomena responsible for the gas and metals exchange among the cold galactic disk, hot galactic envelope, and gas outside of the galaxy halo. In brief, star-formation reheats part of the cold gas in the disk through stellar winds and SNe explosions. Part of this reheated gas remains bound to the host dark-matter halo in a hot gas reservoir associated with galaxies, while a fraction with higher kinetic energy might also escape, and be reincorporated later in the main halo. % Central AGN activity, on the other hand, limits additional cold gas infall toward the galactic disk. 
Overall, a reheating rate, an ejection rate and a re-incorporation rate, fully characterize the feedback scheme of the model. 
We note that, when galaxies become satellites, they are istantaneously stripped of their hot and ejected gas reservoirs, preventing further gas reincorporation (and further gas infall) at later times after the first SF episode. Reincorporation is thus allowed only onto a central galaxy, while satellites continue to consume gas contained in their cold disk without replenishment.

\subsubsection{The semi-analytic model of De Lucia \& Blaizot}\label{deluciablaizot}

\begin{figure*}[ht!]
    \centering
    \includegraphics[angle=0,width=0.95\linewidth,trim={0.cm 7.cm 0.cm 0cm},clip]{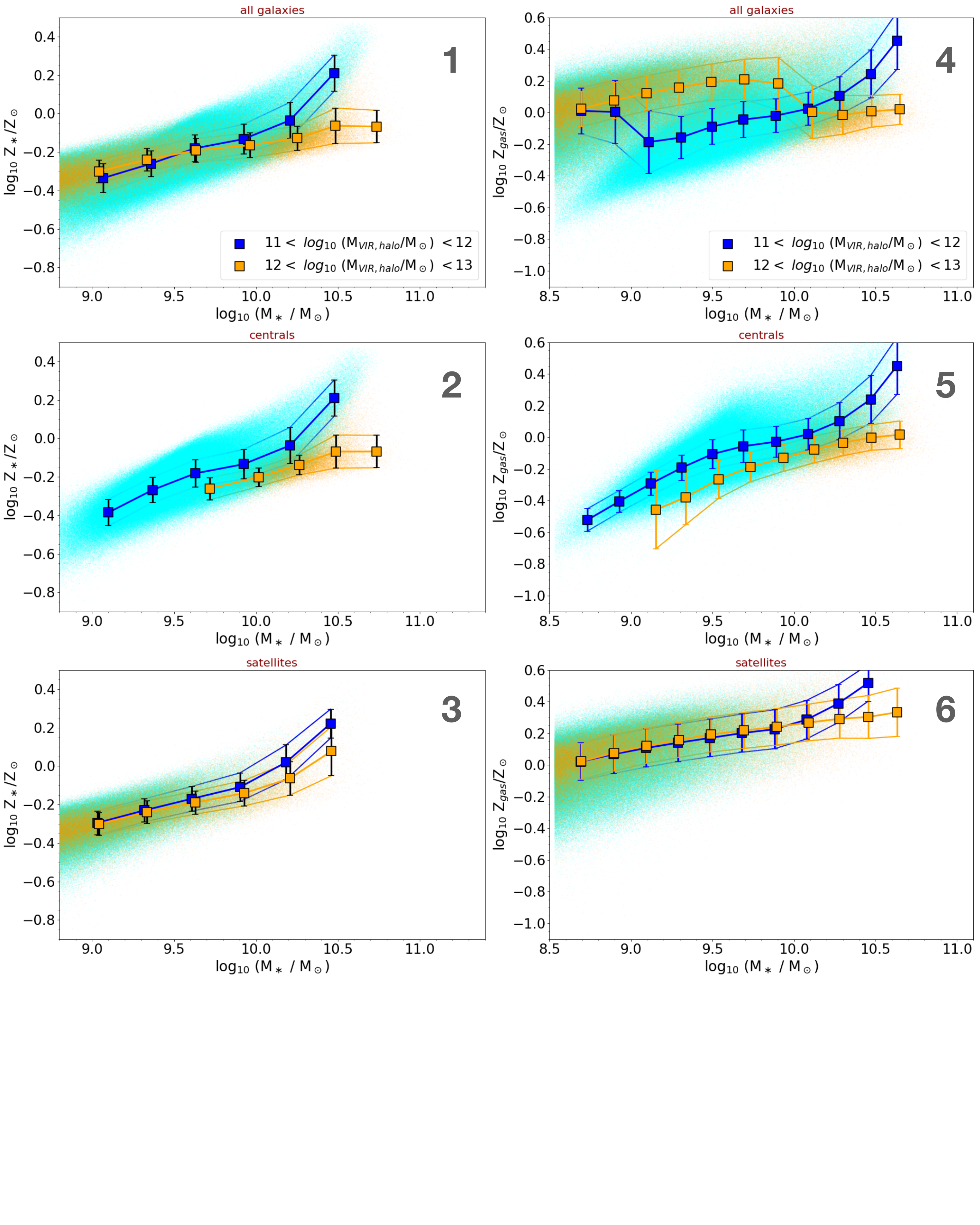}
    \caption{\small Stellar mass - stellar metallicity relation for star-forming galaxies at $z\sim3$ in different bins of virial mass of the host haloes, as predicted by the semi-analytic models of \citet{delucia07}. The relations are drawn for the whole population (panel $3$), for central galaxies only (panel $2$), and for satellites (panel $3$). The vertical error bars represent the median absolute deviation (MAD) of the galaxy metallicities in each stellar mass bin. Panels $4$ to $6$: Same as the plots on the left, but using the metallicity of the cold gas reservoir.}\label{comparison_DeLucia_Blaizot}
\end{figure*}

\begin{figure*}[ht!]
    \centering
    \includegraphics[angle=0,width=0.95\linewidth,trim={0.cm 4.5cm 0.cm 3cm},clip]{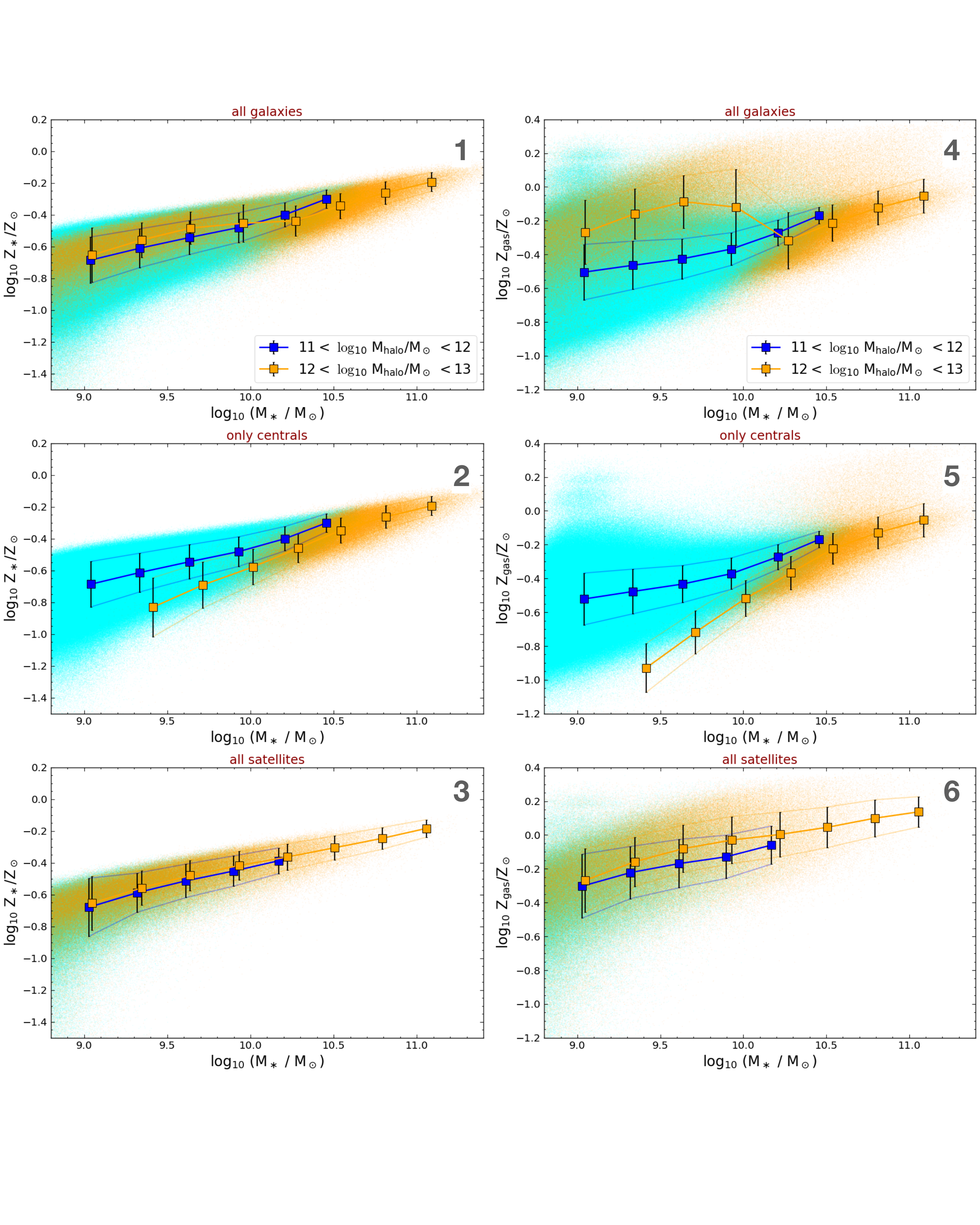}
    \caption{\small Stellar mass - stellar metallicity relation of star-forming galaxies at $z\sim3$ in GAEA, for the whole population (panel $1$), centrals (panel $2$), and satellites (panel $3$), in different bins of virial mass of their host FoF groups from $10^{11}$ to $10^{13}$ M$_\odot$. The vertical error bars represent the median absolute deviation (MAD) of the galaxy metallicities in each stellar mass bin. Panels $4$ to $6$: Same as the left-side panels, but for the cold gas metallicity.}\label{comparison_GAEA}
\end{figure*}

% The DLB07 is the first semi-analytic model that we consider with more detail. 
In this model, the metal enrichment is modeled using the so-called instantaneous recycling approximation (IRA), that is, the total fraction of stellar mass in the form of newly formed metals is returned instaneously to the cold gas disk, neglecting the different evolutionary timescale of stars of different mass. 
We note that this model is able to reproduce both the observed mass-metallicity relation at $z=0$ and the varying baryon fraction between rich clusters and galaxy groups, according to DLB07. 

We consider for our analysis all the galaxies in the model in the redshift snapshot at $z=3$ (close to the median redshift of our work), and we impose SSFR $>10^{-10}$ Gyr$^{-1}$ to reproduce a VANDELS-like selection. 
We then derive for these sample the stellar mass - stellar metallicity relation, which is shown in Fig. \ref{comparison_DeLucia_Blaizot}. 
In the first panel, which comprises all the systems, we can see that galaxies in lower mass haloes (M$_\text{vir}$ ranging $10^{11}$-$10^{12}$ M$_\odot$, identified as blue squares) span similar ranges of stellar masses, from $\sim 10^9$ to $10^{10.6}$ M$_\odot$, of galaxies in higher mass haloes (orange squares).
In the overlapping regions of these two subsets, the metallicities in all the halo mass bins are consistent within $1 \sigma$ dispersion in the stellar mass range between $10^{9}$ and $10^{10.3}$ M$_\odot$, with no evident systematic offsets. Only at the highest stellar masses above $10^{10.3}$ M$_\odot$, we notice a diverging trend, with galaxies in more massive haloes that become significantly more metal poor than systems in less massive structures. 

It is useful to investigate also the behaviour for central and satellite galaxies separately, which are shown in the second and third panels of Fig. \ref{comparison_DeLucia_Blaizot}. 
For central galaxies, we find that in the overlapping regions in the MZ plane of our two halo mass bins, galaxies residing in more massive structures have a systematically lower stellar metallicity by $\sim0.1$ dex on average, increasing to $0.15$ dex in both the low-mass and high-mass end of the diagram. This difference is larger than the standard deviation of the metallicity values in each stellar mass bin, indicating a significant environmental effect. We also notice that this difference is of the same order of the typical uncertainties of our observational estimates of Z$_\star$ (C21). 
In contrast, when considering satellite systems, most of the environmental signatures disappear, and the MZRs in all the halo mass bins become consistent within $1 \sigma$ dispersion in the stellar mass range between $10^{9}$ and $10^{10.3}$ M$_\odot$, as for the global population. Only an increasingly larger offset as a function of halo mass appears at higher stellar masses M$_\star$ $>10^{10.3}$ M$_\odot$. 

In the last three panels on the right of Fig. \ref{comparison_DeLucia_Blaizot} we display the same plots for the cold gas reservoir. 
For central systems (panel $5$), we observe a similar segregation of the MZRs as a function of the host halo mass, but now the differences in metallicity are larger, reaching $0.4$ dex between the least and most massive haloes in the overlapping stellar mass ranges ($10^{9}<$ M$_\star$/M$_\odot$ $<10^{10}$). Satellite systems do not show a significant environmental dependence on the host halo mass, as for the stellar MZR. We notice instead that satellites have in general a higher gas-phase metallicity than equally massive centrals by $\sim0.2$-$0.3$ dex, depending on the galaxy and halo mass regime. For these reason, the jump of the MZRs for the global star-forming population (panel $4$), which puts all the objects together, reflects this double behaviour of satellite and central galaxies. 
We also report that similar results to those obtained for the cold gas metallicity are also found when considering the hot gas component, which is present in central galaxies only, with the same metallicity offsets among different halo masses. 

Overall, the predictions of the DLB07 model can be interpreted in terms of equilibrium condition reached inside galaxies among cooling flows, gas reincorporation, and feedback. %, which act selectively in different dark-matter mass haloes. 
For systems in more massive haloes, the first two mechanisms contribute more strongly to decreasing both their gas and stellar metallicity compared to less massive haloes members. 
The stronger environmental effect found for Z$_\text{gas}$ can be easily explained by the fact that metal enrichment, and in general modifications of the metal content in stars, respond more slowly to variations of the gas-phase metallicity of the gas clouds from which they form. Finally, the different behaviour between satellites and equally massive centrals likely originates from their different treatment, as described above. In particular, the absence of further gas reincorporation and cooling inflows for satellites might lead to more metal rich systems at redshift $\sim3$ compared to central counterparts, and more homogeneous metallicity properties as a function of the environment. % Ricorda che questo modello produce satelliti che sono troppo rosse gia' ad alto redshift. 

\subsubsection{The full GAEA model}\label{GAEA}

\begin{figure*}[ht!]
    \centering
    \includegraphics[angle=0,width=0.95\linewidth,trim={0.cm 13.cm 0.cm 0cm},clip]{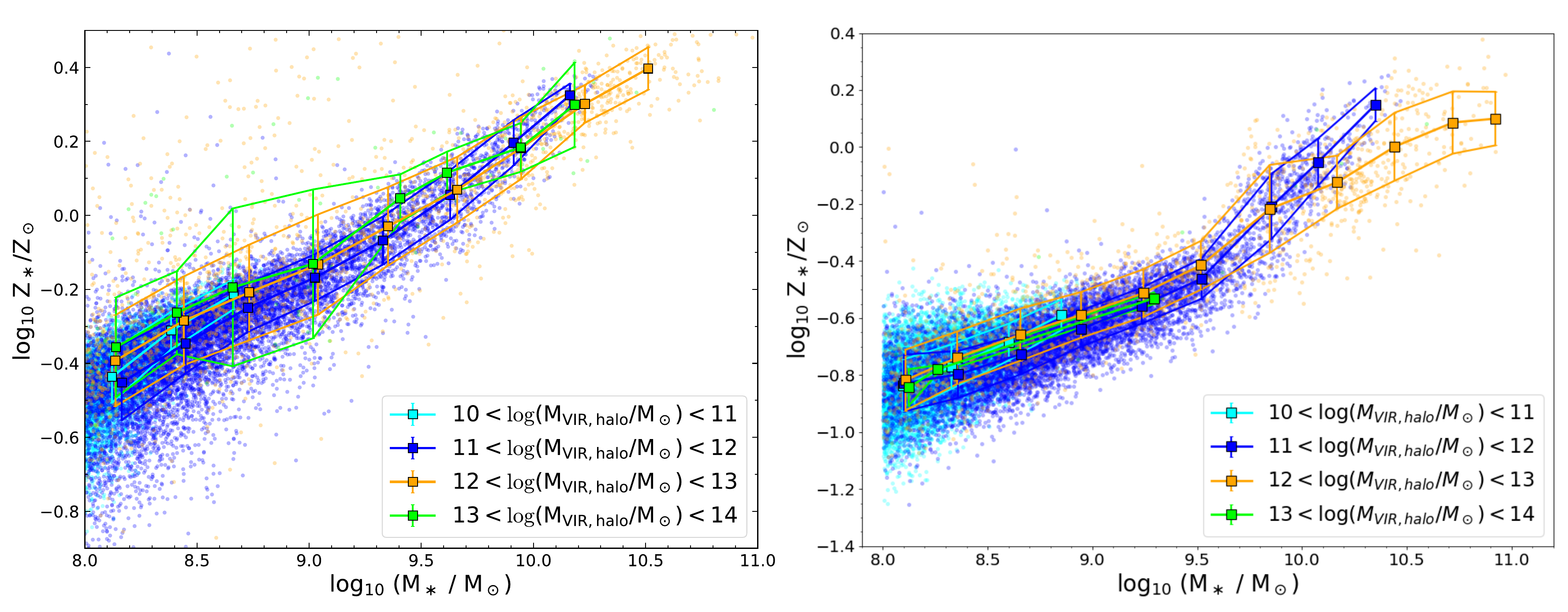}
    \caption{\small \textit{Left}: Stellar mass - stellar metallicity relation of star-forming galaxies at $z \sim 3$ selected from the Eagle hydrodynamical simulation sets (Ref. L0100N1504), as a function of the total mass of their host haloes from $10^{10}$ to $10^{14}$ M$_\odot$. \textit{Right}: Same as the left panel, but for the Illustris-TNG100 simulation.}\label{Eagle_TNG}
\end{figure*}

We also compare our results to the predictions of GAEA \citep{delucia14,hirschmann16,fontanot17}. Since DLB07 and GAEA share the same Millennium cosmological simulation to start with, we only highlight here the main differences between the two models. 
First, GAEA adopts a metal enrichment scheme that explicitely considers the finite lifetimes of stars \citep{delucia14}, thus relaxing the istantaneous recyclying approximation.
The second and most important modification is related to the feedback prescription. The GAEA realization we consider is the same model proposed in \citet{hirschmann16}, which implements a feedback scheme inspired to the results of cosmological zoom-in simulations \citep[the FIRE - Feedback in Realistic Enviroments - project,][]{hopkins14}.
Compared to the previous model, the mass-loading of reheated and ejected gas in GAEA (i.e., the amount of gas reheated and ejected per unit solar mass formed) is more than a factor of two higher at all stellar masses and all redshifts $\gtrsim 2$. Another important ingredient in GAEA, which differs from DLB07 prescription, lies in the treatment of the timescale for reincorporation. This physical process, which brings fresh gas back onto the galactic disk, is not constant in GAEA, but is inversely proportional to the halo mass. In addition, gas cooling is completely suppressed below a virial temperature of $10^4$ K, %while in the original model low-mass haloes were able to cool the same amount of gas of a $10^4$ halo with similar metallicity. 
leading to a delay of gas reaccretion to progressively lower redshifts for lower mass systems, which mimicks the `anti-hierarchical' galaxy evolution scenario. %, and avoiding the rapid formation of very red dwarf satellites too early in time, which was a problem in the previous model.  
Thanks to its improved modelling, GAEA is able to match the stellar mass function at $z=0$ and reproduce its observed evolution at higher redshifts up to at least $z\simeq7$ \citep{hirschmann16,fontanot17}. GAEA models are also able to reproduce the observed evolution of the gas-phase MZR up to $z\sim3.5$ \citep{fontanot21}. Regarding the stellar-phase MZR, it is able to correctly predict the observed slope up to redshift $\sim4$ (C21) and the absolute normalization up to $z < 0.7$ \citep{fontanot21}. 

In order to analyze environmental effects, we consider galaxies selected from the full GAEA model in a small redshift interval around $z=3$ ($\Delta z=\pm0.1$) and with a VANDELS-like requirement on the specific SFR (SSFR $> 10^{-10}$ Gyr$^{-1}$), and divided them in two bins according to their host FoF halo mass from $10^{11}$ to $10^{13}$ M$_\odot$. We then derive the MZR for each subset, which is shown in Fig. \ref{comparison_GAEA}. 
Similarly to the previous model, no environmental effects are observed in the stellar metallicity - stellar mass plane for the global star-forming galaxy population and for satellite galaxies (panels $1$ and $3$). 
For central galaxies in the second panel we see that systems residing in more massive haloes with M$_\text{halo}>10^{12}$ M$_\odot$ (orange squares) show a lower metallicity compared to equally massive galaxies in haloes ranging $11<\log_{10}$ M$_\text{halo}$/M$_\odot$ $12$ (blue squares) in the whole overlapping stellar mass regime between $10^{9.5}$ and $10^{10.5}$ M$_\odot$, but the difference is on average larger ($\sim0.15$ dex) compared to DLB07. This discrepancy is increasing toward lower stellar masses, reaching $0.2$ dex in the bin corresponding to M$_\star \sim 10^{9.5}$ M$_\odot$. 

We then consider the metallicity in the cold gas component in the right panels of Fig. \ref{comparison_GAEA}. The gas-phase MZR of star-forming galaxies in GAEA displays a complex behavior and a much larger scatter than the stellar related diagram. Galaxies in more massive haloes are on average more metal rich compared to systems in less massive structures (panel $4$).  
However, this reflects mostly the offset found between satellite and central galaxies, with the former dominating the full sample at M$_\star < 10^{10}$ M$_\odot$. This difference of cold gas metallicity between the two types of systems can be understood by taking into account the mechanism of complete removal of the hot and ejected gas reservoirs in satellites. This also prevents further accretion of fresh, metal poor gas, a process known to as `strangulation', leading to a gradual increase of metals with respect to central counterparts. 
As a result, the scatter of gas metallicity - stellar mass relation is driven both by the host halo mass and by the relative abundance of satellites and central systems.
For satellite galaxies only (panel $6$), those in the most massive halos are more metal rich than in less massive ones, even though by only $0.1$ dex or less. This is at variance with our findings and to the prediction of the previous model.
Central galaxies instead behave like in the DLB07 model, with more massive haloes (M$_\text{halo} > 10^{12}$ M$_\odot$, orange squares) hosting more metal-poor galaxies than in lower mass haloes (blue squares), and the differences increase up to $0.5$ dex toward lower stellar masses.  

It is important to note that the different criteria adopted for the environmental definition in our semi-analytic models, either by selecting different host halo masses or binning in local density and protocluster membership (by using the same algorithm adopted for the observations on the mock galaxy catalog), produce the same results. We also remark that, even though a metallicity difference is predicted by models only for the central galaxy population, we do not have any method to evaluate if our galaxies are centrals or satellites, hence we postpone this detailed comparison to future studies. 

\subsubsection{Comparison to hydrodynamical simulations}\label{hydrodynamical}

Finally, we briefly mention that we also checked the predictions of hydrodynamical simulations, in particular of Eagle and Illustris-TNG, which are among the most used in the literature. % with a large volume probed. 
The EAGLE project is a large-scale hydrodynamical simulation in a Lambda-Cold Dark Matter universe, run by the Virgo Consortium to study galaxy formation and evolution, and their exchange of gas with the environment \citep{schaye15,crain15}.  
This simulation follows both the dark matter and the baryonic components with a minimum particle mass of $9.7 \times 10^6$ and $1.81 \times 10^6$ M$_\odot$, respectively, yielding a physical picture of galaxy formation from cold gas infall, feedback activity from stellar winds, supernovae explosions, and supermassive black-holes. The efficiency of the feedback processes is calibrated to the present ($z=0$) observed galaxy stellar mass function, the relation between galaxy and central black hole mass, and considers the observed size of galaxies. In our work, we take the public release of halo and galaxy catalogues by \citet{mcalpine16}, and we consider the reference run called L0100N1504, which has the largest volume available with a box size of $100$ Mpc. 

On the other hand, the IllustrisTNG project includes a series of cosmological magnetohydrodynamical simulations of galaxy formation, using state of the art numerical codes and among the most performing supercomputers \citep{weinberger17,pillepich18a}. 
The TNG model includes all the key physical ingredients that can explain different aspects of galaxy properties at varying cosmic times. The chemical enrichment is given by supernovae Ia, II, and AGB stars, and it tracks single elements individually. %The AGN feedback can be either of thermal 'quasar' mode at high accretion rates or of kinetic 'wind' mode when the central black hole is less active. 
The evolution of the cosmic magnetic field is also taken into account. 
These models were shown to successfully reproduce the color-magnitude diagram of galaxies in the local Universe, matching very well the observed distribution from the Sloan Digital Sky Survey and explaining also the physical differences (i.e., SFR, gas fractions and gas metallicities) between the red sequence and the blue cloud galaxy populations \citep{nelson18}. 
Another work by \citet{pillepich18b} focuses instead on the stellar mass distribution inside galaxy groups and clusters, showing that TNG100 well reproduces the observed galaxy stellar mass functions in clusters at $z<1$ for M$_\star < 10^{11}$ M$_\odot$.
We adopt in this paper the public dataset of TNG100 \citep{nelson19}, which simulates a cosmic box volume of $100$ Mpc size and has a resolution of $1.4 \times 10^6$ M$_\odot$ in baryon mass and $7.5 \times 10^6$ M$_\odot$ in dark matter mass. %, similar to the original Illustris runs. 
We note that this choice was taken to reach the best compromise between large area and high resolution among the different TNG runs, even though the final conclusions would not change significantly if we consider a different set of simulations with a different box size and resolution.
In both TNG and EAGLE, we consider star-forming galaxies with SSFR $>10^{-10}$ Gyr$^{-1}$ to mimick the VANDELS selection.

We display as representative the MZR obtained with these two hydrodynamical codes for the global star-forming galaxy population, which can be directly compared to the entire star-forming sample studied in VANDELS (Sect. \ref{MZRenvironment}). In these cases, we explore the environmental dependences in a larger range of halo masses from from $10^{10}$ to $10^{14}$ M$_\odot$, %These two plots are the most representative results of Eagle and TNG, as they can be directly compared to the whole observational sample studied in Sect. \ref{MZRenvironment}. 
%As before, we present the stellar mass - stellar metallicity relation as a function of different halo masses, from $10^{10}$ to $10^{14}$ M$_\odot$, 
where the haloes are always identified through a standard FoF algorithm, as in semi-analytic models. % with linking length of $0.2$ (check this). 
As we can see in Fig. \ref{Eagle_TNG}, in the overlapping regions within the stellar mass range from $10^{8.5}$ to $10^{10.5}$ M$_\odot$, which is that probed by our VANDELS data, we find no significant offset between galaxies residing in less and more massive haloes, with metallicity differences in each bin lower than $0.05$ dex, much lower than the statistical uncertainties. We only observe a slightly larger dispersion of the MZR for the more massive haloes population in the Eagle run. 
The same qualitative result of no significant environmental dependence is found in our stellar mass range when considering the gas-phase metallicity in both simulations.

\subsubsection{Final considerations on the environmental modeling at $z\sim3$}\label{final}

The models and simulations presented in the previous subsection, namely the GAEA and \citet{delucia07} SAMs, Eagle and Illustris-TNG, all agree qualitatively on the absence of significant environmental dependence of the MZR for the star-forming galaxy population. %, using a multitude of environmental diagnostics. 
We have also learned that the exact modeling of the chemical enrichment and feedback prescriptions have a small influence on this qualitative result. 
Even though more statistics is needed to better constrain the significance of our observational finding and in particular the gas-phase MZR in a larger range of stellar mass, our finding suggests that one would need to account explicitly for environmental dependent processes in the models (like those discussed in Section \ref{discussion1}), which are currently not included. 

We have also seen in SAMs that central galaxies living in more massive haloes have a lower stellar metallicity compared to objects in less massive structures, which resembles some of the VANDELS observational results presented in Section \ref{results}, showing a weak decrease of the metallicity when we move toward denser regions and protoclusters. 
A possible physical interpretation is that the overdensity code adopted in G20 connects structures larger than the typical virial radius of dark-matter haloes at these redshifts, thus connecting different independent haloes at slightly different redshifts and sky positions (each with its own central galaxy), and we might have a larger fraction of galaxies close to the overdensity peaks that behave like centrals. 

Furthermore, all the models (both GAEA and DLB07) converge toward a systematic difference in metallicity between central and satellite galaxies, with the latter having on average a higher metal content. In the local Universe, stars in SF satellites are more metal-rich than equally massive centrals (Pasquali et al. 2010), and this seems to be in place also at redshift $\sim3$ from a theoretical point of view. 
However, we remind that a proper comparison between models and observations for central galaxies and satellites separately is not a goal of this paper, and will represent a challenge for future works. % exploiting the potential of new observational facilities. 
Testing observationally the metallicity offset between centrals and satellites remains difficult, and would need higher statistics and depth at these redshifts, which can be achieved in future large field surveys like Euclid, LSST, WFIRST, GMT, and TMT.

\section{Summary and conclusions}\label{conclusions}

In this paper we have investigated the effects of the large scale environment on the relation between stellar mass and both the stellar and gas-phase metallicity, for star-forming galaxies in the redshift range $2<z<4$ selected from the VANDELS spectroscopic survey. We have tested different criteria for defining the environment in which galaxies live. These are based primarily on the local density map calculated from the VANDELS parent sample covering the entire cosmic volume probed by the survey, and afterwards on the overdensity structures and protoclusters identified from those maps. 
The main findings are summarized in the following:

\begin{itemize}
\item We analyze the stellar phase mass-metallicity relation (MZR) as a function of local galaxy density. We find that the MZR does not depend significantly on the density of galaxies from M$_\star$ $=10^{9.4}$ to $10^{10.5}$ M$_\odot$, while a metallicity offset at $1\sigma$ significance is present for lower stellar masses, with galaxies in the densest regions ($\rho_{gal} \simeq 0.01$ gal/Mpc$^3$) being less metal-rich by $\sim0.3$ dex than more isolated analogs in underdense areas and by $\sim0.2$ dex compared to the global MZR.
\item Galaxies inside overdensity structures and protoclusters with different radii have a lower stellar metallicity compared to galaxies in the field by $0.2$ dex on average (significance of $1\sigma$), indicating that the environmental effects are more important in the proximity of the densest cores and protocluster centers.
\item We analyze the environmental effects on the gas-phase MZR for a smaller subset of VANDELS galaxies with complementary near-infrared spectra. We find a weak trend suggesting that galaxies in denser regions have more metal poor gas than similarly massive systems in underdense regions. Even though this analysis is limited by the low statistics, our trend corroborates recent findings based on a larger statistical sample that suggest a reversal of the gas metallicity-environmental dependence at $z>2$ \citep{chartab21}. 
\item The slightly lower metallicity inside dense cores and protoclusters is not driven by selecting galaxies with a higher level of SFR, fainter UV magnitude M$_{1600}$, lower dust attenuation, or younger age compared to the field, but it is likely due to real environmental effects. We suggest as possible interpretations a higher efficiency of gas stripping phenomena and harassment in protoclusters, outflows induced by enhanced merger events or AGN powered, all of which can deprive galaxies from their metal-enriched gas, or an efficient dilution of metals in the cold gas reservoirs driven by pristine gas inflows from the cosmic web at $z>2$, or by galaxy interactions. While these effects might plausibly work in combination, testing the influence of each individual mechanism is out of our capabilities and will be achievable with forthcoming large-scale surveys. 
\item The absence of a significant environmental dependence of the SFR and dust attenuation for main-sequence star-forming galaxies at redshift $2<z<4$ in our mass range ($10^9 < M_\star/M_\odot < 10^{10.5}$) suggests that cold gas accretion from the cosmic web, if ubiquitously present in high-redshift protoclusters, mainly fuels dust-obscured star-formation, as claimed in the recent literature, even though it might be in part stored in the circumgalactic medium of normal galaxies and consumed on much longer timescales at slower rate.
\item Semi-analytic models of galaxy evolution and hydrodynamical simulations predict no significant environmental dependence of the mass-metallicity relation for the global star-forming galaxy population. This permaning tension suggests that environmental dependent processes in overdense regions must be explicitly taken into account in the models. 
\end{itemize}

\noindent
Testing environmental effects on the satellite and central population separately cannot be reached with current data and will be challenging for next generation surveys with Euclid, LSST, MOONS, WFIRST, GMT, and TMT. Thanks to their depth and the larger object statistics achievable, these will allow the derivation of galaxy density maps with good precision up to higher redshifts (from $z=4$ to $\sim7$), and the identification of the first protocluster structures in the Universe. Given the low S/N nature of our results, it will be important in the future to study the environmental effects on the mass - metallicity relation with much more statistical power in larger fields. %we expect to obtain very soon large improvements in statistics with JWST observations.

Finally, including multi-wavelength data to probe additional galaxy properties in our study will be crucial to obtain a more complete picture of the galaxy trasformation that is thought to happen in clusters at around redshift $\sim2$, shedding light on the physical origin of the environmental driven metallicity offset found in this work around the densest protocluster cores at $z>2$.

\begin{acknowledgements}
We thank the referee for thoughtful and constructive comments that have improved the quality of this manuscript. AC acknowledge the support from grant PRIN MIUR 2017 20173ML3WW$\_$001. MLl acknowledges support from the ANID/Scholarship Program/Doctorado Nacional/2019-21191036. %RA acknowledges support from ANID FONDECYT Regular 1202007.
\end{acknowledgements}

\clearpage
\appendix
\section{Dependence of the metallicity on the spectrum S/N}\label{appendix1}

\begin{figure}[h!]
    \centering
    \includegraphics[angle=0,width=\linewidth,trim={0cm 0cm 0cm 0cm},clip]{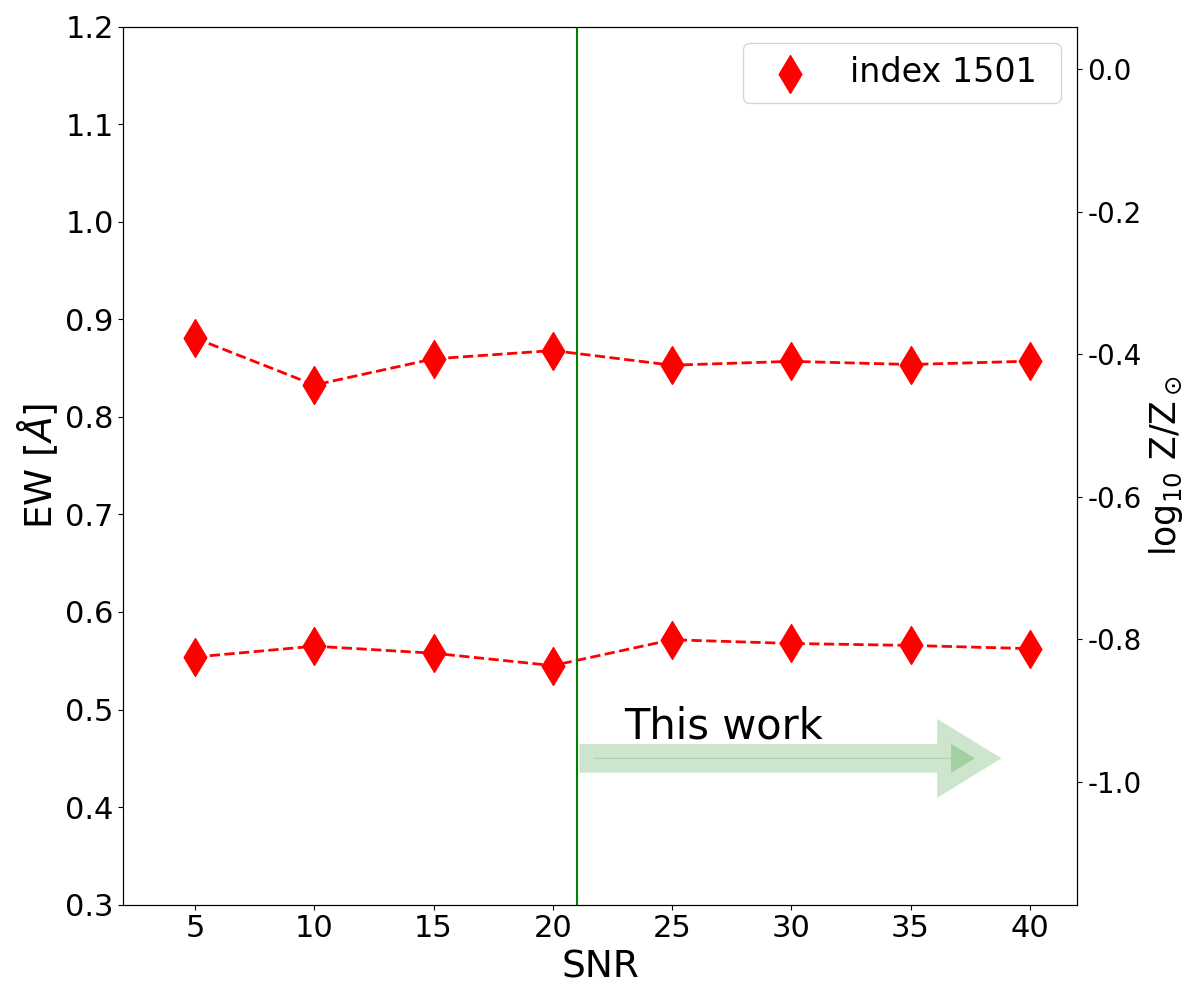} 
    \includegraphics[angle=0,width=\linewidth,trim={0cm 0cm 0cm 0cm},clip]{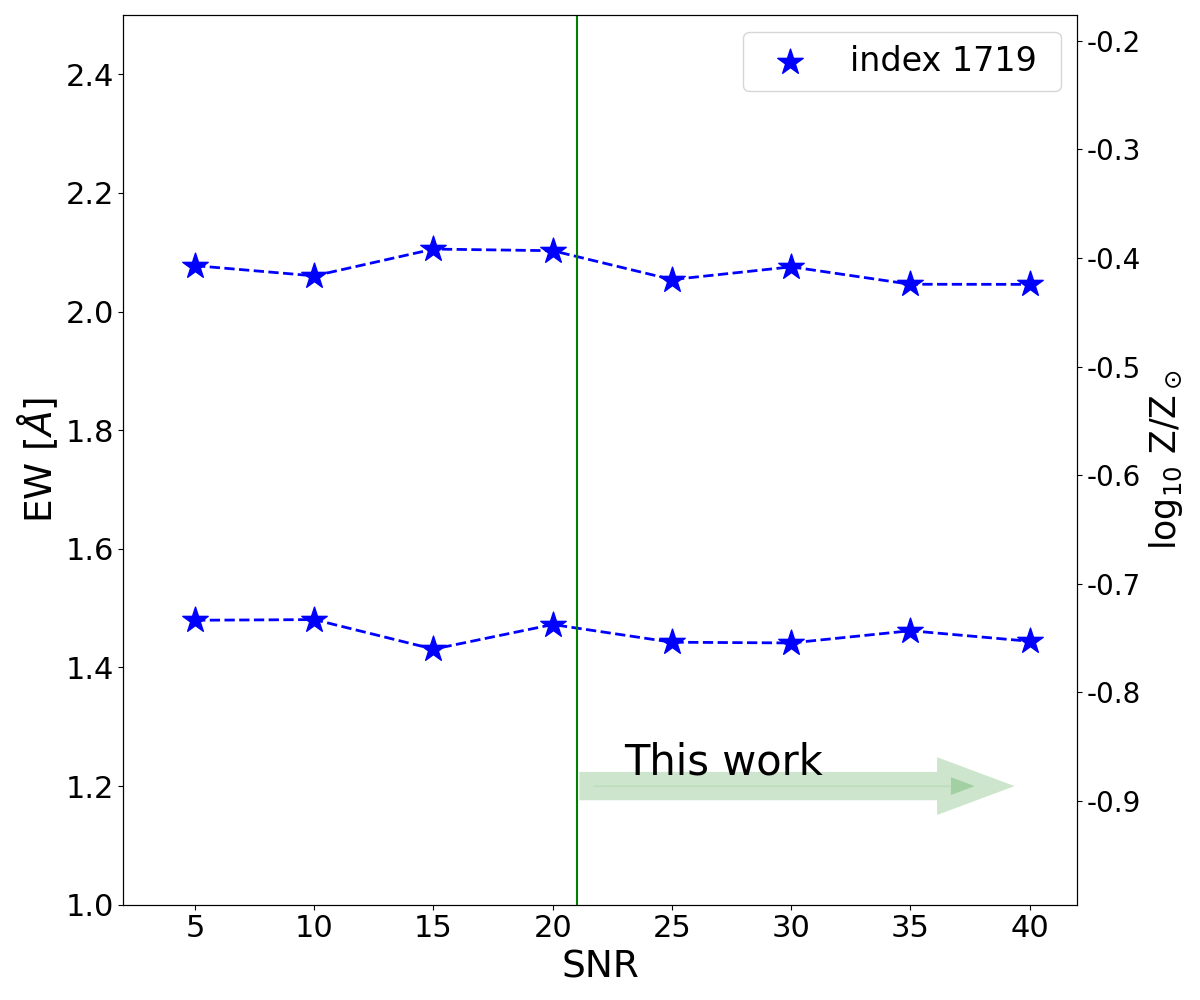}
    \caption{\small \textit{Top:} Relation between the measured EW (and recovered stellar metallicity on the left y-axis, using the calibrations in Equations \ref{calib1719}) and input spectrum S/N for the $1501$ \AA\ index. This is obtained through Monte Carlo simulations where Gaussian noise is added to Starburst99 templates with two different metallicities Z$_\star = 0.2$ and $0.4$ $\times$ Z$_\odot$. The vertical green line highlights the S/N range of the stacks that we analyze in our work. \textit{Bottom:} Same of the upper panel, but for the $1719$ \AA\ index. }\label{Zvalue_SNR}
\end{figure}

Since we have shown in C21 (Figure A.1) that the S/N of the spectra strongly affects the error that we obtain for the equivalent width of the absorption lines (hence for the metallicity), we might wonder whether the S/N of our stacks also plays a role in the estimation of the equivalent width itself, producing a bias in the recovered metallicity value. We test this effect by using the same simulations already adopted in C21, taking two Starburst99 models at different intrinsic metallicity values ($0.2$ and $0.4$ times solar), reporting them to the VANDELS resolution as done for the calibrations, and perturbing the theoretical templates with increasing Gaussian noise, from S/N $=5$ to $40$. We finally take the average EW and metallicity for the $1501$ and $1719$ index out of $500$ realizations. As shown in Figure \ref{Zvalue_SNR}, in all cases there is no dependence of the recovered metallicity on the S/N of the stellar continuum, with fluctuations that are always well below $0.01$ dex in the S/N range of our VANDELS spectra. As a result, we conclude that the small metallicity differences seen in Section \ref{MZRenvironment} are not driven by the different S/N levels reached by the two galaxy populations in the field and in overdensities. In this regard, we notice that a difference in metallicity is also found when we adopt the largest definition of protoclusters (Fig. \ref{MZR_environment2}- panel $4$), where the measurement S/N of the two galaxy populations inside and outside of the structures are similar. 

\end{document}